\documentclass[fleqn,usenatbib]{mnras}

\usepackage{newtxtext,newtxmath}

\usepackage[T1]{fontenc}

\DeclareRobustCommand{\VAN}[3]{#2}
\let\VANthebibliography\thebibliography
\def\thebibliography{\DeclareRobustCommand{\VAN}[3]{##3}\VANthebibliography}

\usepackage[pdftex]{graphicx}	
\usepackage{graphicx}	
\usepackage{amsmath}	
\usepackage{tabto}

\title[Gas-phase formation of acetaldehyde]{Gas-phase formation of acetaldehyde: review and new theoretical computations}

\author[F. Vazart et al.]{
Fanny Vazart,$^{1}$\thanks{E-mail: fanny.vazart@univ-grenoble-alpes.fr}
Cecilia Ceccarelli,$^{1}$
Nadia Balucani,$^{2}$
Eleonora Bianchi$^{1}$
\newauthor and 
Dimitrios Skouteris$^{3}$
\\
$^{1}$Univ. Grenoble Alpes, CNRS, Institut de Planetologie et d'Astrophysique de Grenoble (IPAG), 38000 Grenoble, France\\
$^{2}$Dipartimento di Chimica, Biologia e Biotecnologie, Università degli Studi di Perugia, 06123 Perugia, Italy\\
$^{3}$MASTER UP, 06123 Perugia, Italy\\
}

\date{Accepted XXX. Received YYY; in original form ZZZ}

\pubyear{2020}

\begin{document}
\label{firstpage}
\pagerange{\pageref{firstpage}--\pageref{lastpage}}
\maketitle

\begin{abstract}
Among all the interstellar complex organic molecules (iCOMs), acetaldehyde is one of the most widely detected species.
The question of its formation route(s) is, therefore, of a major interest regarding astrochemical models. 
In this paper, we provide an extensive review of the gas-phase formation paths that were, or are, reported in the literature and the major astrochemical databases. 
Four different gas-phase formation routes stand out : \textit{(1)} CH$_3$OCH$_3$ + H$^+$ / CH$_3$CHOH$^+$ + e$^-$, \textit{(2)} C$_2$H$_5$ + O($^3$P), \textit{(3)} CH$_3$OH + CH and \textit{(4)} CH$_3$CH$_2$OH + OH / CH$_3$CHOH + O($^3$P). 
Paths \textit{(2)} and \textit{(3)} were not studied neither via laboratory or theoretical works in the low temperature and density regime valid for the ISM. 
Thus, we carried out new accurate quantum chemistry computations. 
A theoretical kinetics study at low temperatures (7$\div$300 K), adopting the RRKM scheme, was also performed.
We confirm that reaction \textit{(2)} is efficient in forming acetaldehyde in the 7-300 temperature range ($\alpha$ = $1.21\times10^{-10}$ cm$^3$ s$^{-1}$ and $\beta$=0.16). 
On the contrary, our new computations disprove the formation of acetaldehyde through reaction \textit{(3)} ($\alpha$ = 1.84$\div$0.67$\times10^{-13}$ cm$^3$ s$^{-1}$ and $\beta$=-0.07$\div$-0.95). 
Path \textit{(1)} was showed to be inefficient too by recent computations, while path \textit{(4)} was formerly considered for glycolaldehyde formation, having acetaldehyde as a by-product. 
In conclusions, of the four above paths only two, the \textit{(2)} and \textit{(4)}, are potentially efficient gas-phase reaction routes for the formation of acetaldehyde and we encourage astrochemical modellers to only consider them. 
Comparison with astronomical observations suggest that path \textit{(4)} may actually play the major role. 
\end{abstract}

\begin{keywords}
astrochemistry < Physical Data and Processes
\end{keywords}


\section{Introduction}

The formation of interstellar complex organic molecules \citep[iCOMs;][]{ceccarelli2017}, namely organic species composed of more than five atoms \citep{Herbst2009}, is of particular importance in astrochemistry. 
Indeed, iCOMs are detected in various regions, such as star-forming regions \citep{Rubin1971,cazaux2003,Kahane2013,Mendoza2014,belloche2017,ligterink2017,Mcguire2018}, circumstellar envelopes of AGB stars \citep{cernicharo2000} or shocked regions \citep{Arce2008,codella2017,lefloch2017}. 
Understanding how these quite "complex" compounds could be formed in the harsh conditions of ISM, \textit{i. e.} at very low temperatures and densities, is a challenging question. 
So far, two main, and not incompatible, chemical theories have been invoked: solid-state chemistry \citep{Garrod2006,Woods2013,fedosev2015,oberg2016} and gas-phase reactivity \citep{Charnley1992,Balucani2015,Skouteris2018}. 

In this article we focus on acetaldehyde (CH$_3$HCO), one of the first molecules to have been detected in the interstellar medium (ISM) \citep{Gottlieb1973} and one of the most abundant iCOMs.
Indeed, acetaldehyde is almost ubiquitously detected, in cold ($\sim 10$ K) and warm ($\geq 50$ K) environments \citep[e.g.]{Blake1987, cazaux2003, Bacmann2012, Vastel2014, lefloch2017, Sakai2018, Bianchi2019, Lee2019, Csengeri2019, Scibelli2020, DeSimone2020}.
Furthermore, acetaldehyde has a great prebiotic potential, being a possible precursor for several carbohydrates \citep{Pizzarello2004,cordova2005} and acrolein (CH$_2$CHCHO).
\begin{figure*}
	\includegraphics[scale=0.47]{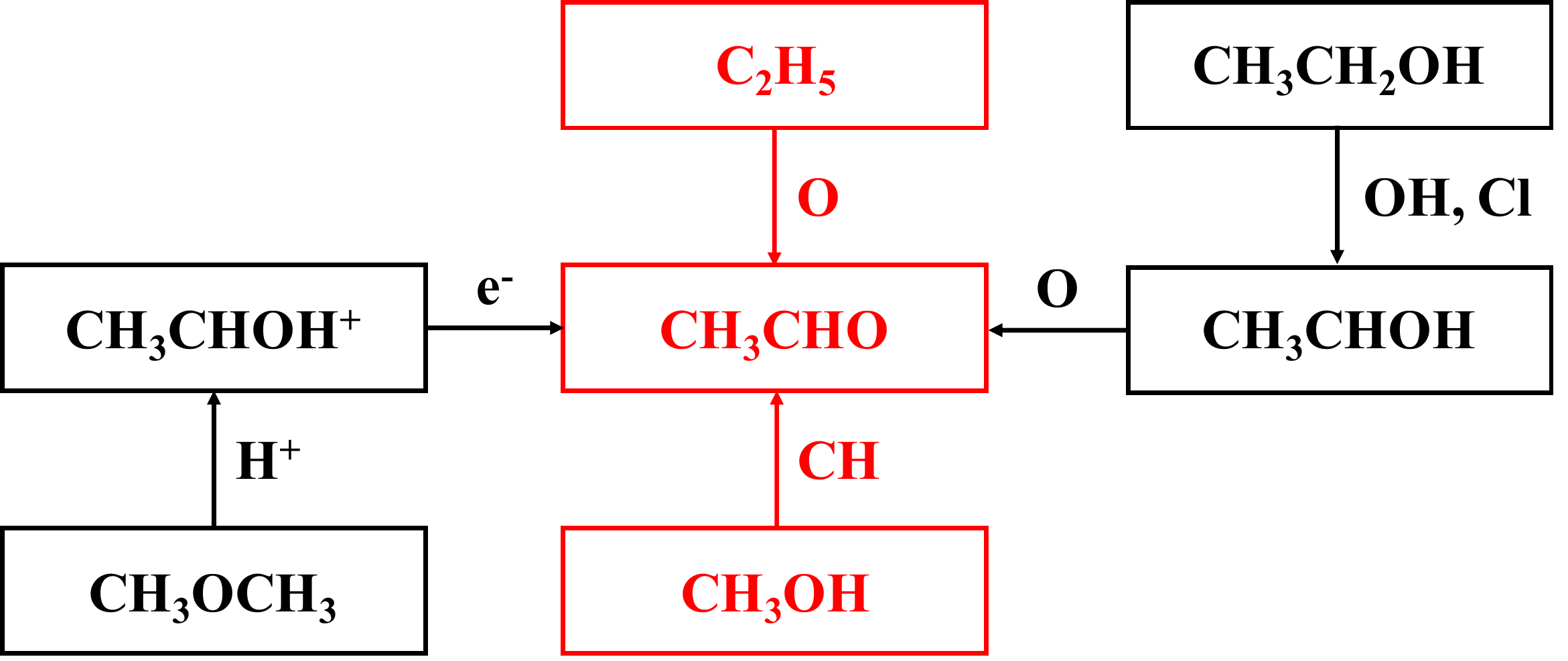}
    \caption{Scheme of the four gas-phase formation routes of acetaldehyde according to the reactions proposed in the literature and listed in \S ~\ref{sec:reviews}. 
    The boxes in red mark the two reactions studied in the present work.}
    \label{fig:scheme-react}
\end{figure*}
The latter is a crucial intermediate in the prebiotic synthesis of various amino acids \citep{cleaves2003}. 
It can also be considered as a condensation agent in the prebiotic formation of deoxyribonucleosides \citep{teichert2019}, a major component of DNA, and was used by Adolph Strecker in 1850 \citep{strecker1850} in his famous amino acid synthesis to form alanine.  

Despite the ubiquity of acetaldehyde in the molecular ISM and its potential prebiotic importance, there is still not a consensus of how this molecule is formed.
It could be the product of the chemistry occurring on the grain ice surfaces \citep[e.g.][]{Garrod2006,Jones2011,Bennett2005,Martin-Domenech2020} or synthesised in the gas-phase \citep[e.g.][]{charnley2004,Vastel2014,Codella2020,DeSimone2020}.
In this article, we focus on the gas-phase formation routes that have been proposed in the literature.
Our aim is to provide a completely validated network of reactions that form acetaldehyde and that can then be used in astrochemical models.

The article is organised as follows.
In Section \ref{sec:reviews}, we provide a summary of the reactions present in the literature.
For the cases where no experimental or theoretical estimates exist in the range of temperatures and pressures valid in the molecular ISM, we carried out new theoretical computations, both on the electronic energy and kinetics.
Section \ref{sec:methods} describes the adopted computational methods and Section \ref{sec:results} the results of the computations.
In Section \ref{sec:discussion}, we discuss the implications of our new computations, and provide guidelines on the reactions and rate constants (please note that astronomers tend to write "rate coefficients", which is an equivalent terminology) to be used in astrochemical models, after the comparison with astronomical observations.
Section \ref{sec:conclusions} concludes the article.

\section{Gas-phase routes to acetaldehyde formation}
\label{sec:reviews}

Several gas-phase acetaldehyde formation reaction paths have been proposed in the literature.
They involve ion-molecule and neutral-neutral reactions. 
Most of them are included in the two major databases used by astrochemical modellers, KIDA \citep[Kinetic Database for Astrochemistry:][]{Wakelam2012} and UDfA \citep[UMIST Database for Astrochemistry:][]{mcelroy2013}.
Among the various reaction paths listed in those two databases two are potentially efficient to synthesise acetaldehyde in the gas-phase: 

\medskip
\begin{tabular}{ll}
     (1) & CH$_3$OCH$_3$ + H$^+$ $\rightarrow$ CH$_3$CHOH$^+$ + H$_2$\\
         & CH$_3$CHOH$^+$ + e$^-$ $\rightarrow$ CH$_3$CHO + H\\
     (2) & C$_2$H$_5$ + O($^3$P) $\rightarrow$ CH$_3$CHO + H\\
\end{tabular}

\medskip
A third path was recently proposed by \cite{Vasyunin2017} and it is also reported in the UDfA database:

\medskip
\begin{tabular}{ll}
     (3) & CH$_3$OH + CH $\rightarrow$ CH$_3$CHO + H\\
\end{tabular}

\medskip
A fourth path that starts from ethanol (CH$_3$CH$_2$OH) was finally proposed by \cite{Skouteris2018}:

\medskip
\begin{tabular}{ll}
     (4) & CH$_3$CH$_2$OH + OH $\rightarrow$ CH$_3$CHOH + H$_2$O\\
         & CH$_3$CHOH + O $\rightarrow$ CH$_3$CHO + OH\\
\end{tabular}

\medskip
The UDfA database also reports the reaction CH$_3$CHCH$_2$ + OH $\rightarrow$ CH$_3$CHO + CH$_3$. 
However, the formation of acetaldehyde from this reaction is very unlikely, as it would require several steps in the reaction path and acetaldehyde would certainly be a (very) minor product. 
We, therefore, do not consider this path further.

\medskip
While paths (1) and (4) were studied by \cite{vazart2019} and \cite{Skouteris2018}, respectively, via theoretical computations of the electronic energy and kinetics of the involved reactions, paths (2) and (3) have not been validated yet neither by experimental or theoretical works in the conditions of ISM, namely low temperatures and pressure.
Specifically, reaction (2) was studied by combined cross-beam and computational studies. However, these studies were not focused on kinetics and they were carried out in the 295-600 K temperature range \citep{Jung2011,Jang2014,Park2010}.
Reaction (3) was studied computationally, using an ab initio and DFT composite method but no kinetic computations were carried out \citep{Zhang2002}. 
It was also studied experimentally in the 298-753 K temperature and 100-600 Torr pressure ranges \citep{Johnson2000}, which, unfortunately, are conditions not directly applicable to the molecular ISM.

In this work, we carry out new computations to obtain the products and rate constants in the 7-300 K temperature range for the reactions (2) and (3), in order to have a complete validated network of reactions forming acetaldehyde.
Figure \ref{fig:scheme-react} schematically summarises the four possible routes that leads to the formation of acetaldehyde in the gas-phase and the two that are studied here are pictured in red.

\section{Computational details and methods}
\label{sec:methods}

\subsection{Electronic structure computations and vibrational evaluation}
\label{subsec:comput-methods} 

All the computations were carried out using the Gaussian16 suite of programs \citep{gaussian16}. The B2PLYP double hybrid functional \citep{B2PLYP} was used for all the geometry optimizations, in conjunction to the aug-cc-pVTZ triple-$\zeta$ basis set \citep{aug-cc-pvtz1,aug-cc-pvtz2}. Semiempirical dispersion effects were also included thanks to the D3BJ model of Grimme \citep{D3BJ}, leading to the so-called B2PLYP-D3/aug-cc-pVTZ level of theory. The frequencies of all the involved compounds were also evaluated using this method, in order to verify that all intermediates were true minima on the potential energy surface (PES), and that all transition states (TSs) exhibited a single imaginary frequency. The electronic energies were then reevaluated using the coupled-cluster singles and doubles approximation augmented by a perturbative treatment of triple excitations (CCSD(T), \cite{CCSDT}) in conjunction to the same basis set. This composite method will be referred to as CCSD(T)//B2PLYP-D3/aug-cc-pVTZ in the present manuscript. 

\subsection{Kinetics study methods}
\label{subsec:kinetics-methods} 

As in previous work \citep{Balucani2012,Leonori2013,Skouteris2015,Vazart2015,Skouteris2018} a combination of capture theory and the Rice-Ramsperger-Kassel-Marcus (RRKM) calculations was used to determine the relevant rate constants and branching ratios. For the first steps (the addition of the O($^3$P) atom to C$_2$H$_5$ and the formation of the initial Van der Waals CH...CH$_3$OH complex regarding the first and second reactions, respectively) capture theory was used. To do so, calculations were performed at various long-range distances of the reactants, and the obtained energies obtained were fitted to a $1/R^6$ functional form (both for the London dispersion forces and the rotating dipole ones). The fitting coefficient (C$_6$) was then used to obtain the capture cross section with the formula $\sigma(E) = \pi \times 3 \times 2^{-2/3} \times (C_6/E)^{1/3}$ (where $E$ is the translational energy), which was itself multiplied by the collision velocity $(2E/m)^{1/2}$ (where $m$ is the reduced mass of the reactants) to get the corresponding capture rate constants together with the maximum total angular momentum $J$ for a successful capture. 
For the subsequent reactions, energy-dependent rate constants were calculated using the RRKM scheme, $J$ being conserved throughout it (for each energy, RRKM calculations are carried out separately for all values of $J$ up to the maximum one permitted). Subsequently, the master equation was solved at all relevant energies for all systems (to consider the overall reaction scheme), Boltzmann averaging was carried out to obtain temperature-dependent rate constants and, finallly, those rate constants were fitted to the form $k(T)=\alpha({\frac{T}{300K}})^\beta$ The values of $\alpha$ and $\beta$ in each case are given in Table \ref{tab:alpha-beta} of the \ref{subsec:kinetics-results} following section.

\section{Results}
\label{sec:results}

\subsection{Electronic structures and reaction paths}
\label{subsec:paths} 

\begin{figure*}[h]
	\includegraphics[scale=0.5]{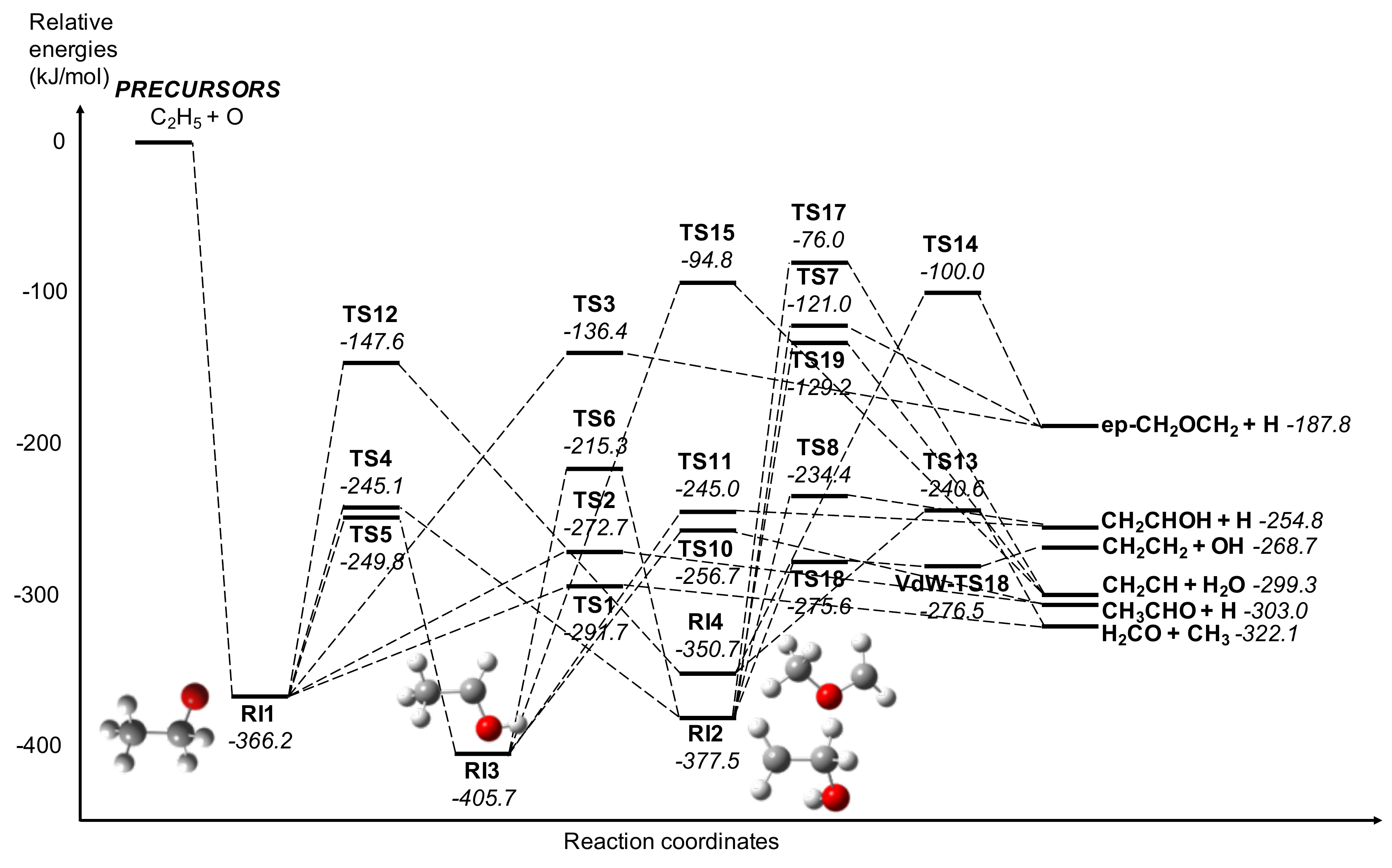}
    \caption{Full reaction path following the addition of O($^3$P) on the C$_2$H$_5$ radical at the CCSD(T)//B2PLYP-D3/aug-cc-pVTZ level of theory. The exhibited energies include the ZPE corrections.}
    \label{fig:path-C2H5+O}
\end{figure*}
\begin{figure*}
	\includegraphics[scale=0.45]{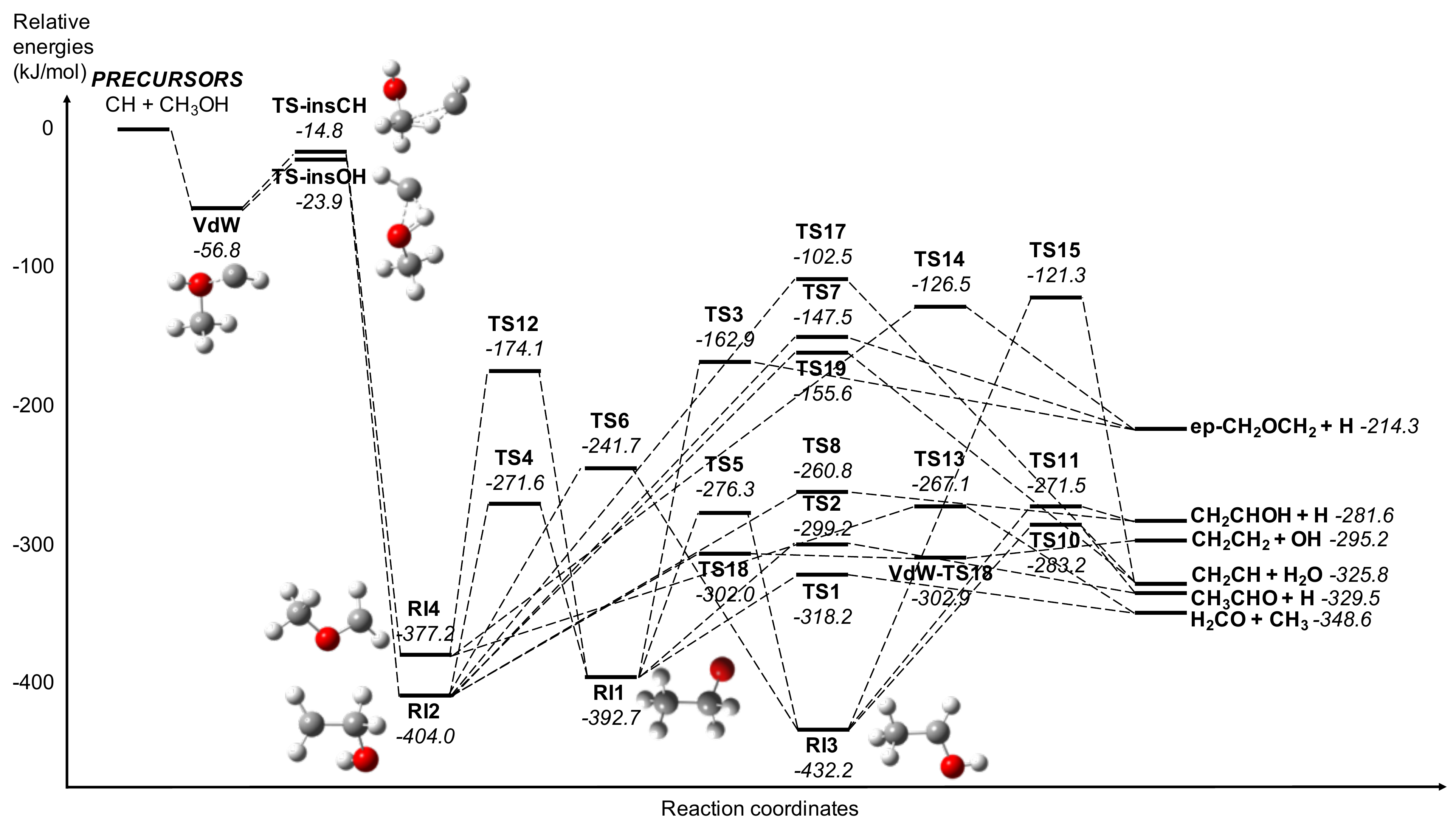}
    \caption{Proposed reaction path of the CH$_3$OH + CH reaction at the CCSD(T)//B2PLYP-D3/aug-cc-pVTZ level of theory. The exhibited energies include the ZPE corrections.}
    \label{fig:path-CH3OH+CH}
\end{figure*}

This section summarizes the electronic structures and relative energies of all relevant intermediates and transition states involved in both C$_2$H$_5$ + O($^3$P) and CH$_3$OH + CH channels. The optimized geometries and energies of each species are given in Appendix.

\textit{C$_2$H$_5$ + O($^3$P).} On Fig. \ref{fig:path-C2H5+O}, the full reaction path following the barrierless addition of O($^3$P) on the radical C$_2$H$_5$ is presented. This path was already proposed by \cite{Jung2011}, at the CBS-QB3 level of theory. The energies shown here are the CCSD(T)/aug-cc-pVTZ reevaluated electronic energies corrected with the B2PLYP-D3/aug-cc-pVTZ zero-point energies (ZPE). Starting from the first \textbf{RI1} intermediate, one can observe three types of direct dissociations: into H$_2$CO + CH$_3$, exhibiting a 75 kJ/mol barrier represented by \textbf{TS1}, into acetaldehyde CH$_3$CHO + H, through \textbf{TS2} which is \textit{ca.} 95 kJ/mol more energetic than \textbf{RI1}, or into the CH$_2$OCH$_2$ epoxide + H, if the system overpasses the 230 kJ/mol large barrier defined by \textbf{TS3}. \textbf{RI1} is also able to be converted into all three other intermediates \textbf{RI2}, \textbf{RI3} and \textbf{RI4}, through the \textbf{TS4}, \textbf{TS5} and \textbf{TS12} transition states, which exhibit 120, 117 and 220 kJ/mol barriers, respectively.

\textbf{RI3}, as far as it is concerned, can also undergo 3 types of direct dissociations: into acetaldehyde CH$_3$CHO + H, through \textbf{TS10} which has a \textit{ca.} 150 kJ/mol barrier, into the CH$_2$CHOH enol + H, after over-passing the 260 kJ/mol barrier represented by \textbf{TS11}, into CH$_2$CH + H$_2$O, a dehydration reaction requiring 310 kJ/mol to occur (Cf. \textbf{TS15}). It is linked to \textbf{RI1} through the previously mentioned \textbf{TS5} and to \textbf{RI2} through \textbf{TS6}, that exhibits a 190 kL/mol barrier.

If we take a look at \textbf{RI2}, linked to \textbf{RI1} and \textbf{RI3} thanks to \textbf{TS4} and \textbf{TS6} respectively, it can undergo five types of dissociations: into CH$_2$CH$_2$ + OH, through \textbf{TS18}, which exhibits a \textit{ca.} 100 kJ/mol barrier, followed by a loose Van der Waals complex \textbf{VdW-TS18}, into the CH$_2$CHOH enol + H, after over-passing the 140 kJ/mol barrier represented by \textbf{TS8} or into the CH$_2$OCH$_2$ epoxide, through the 250 kJ/mol barrier represented by \textbf{TS7}. \textbf{RI2} can also experience two different dehydration steps, over \textbf{TS17} or \textbf{TS19}, that will form CH$_2$CH + H$_2$O after over-passing a 300 or a 250 kJ/mol barrier, respectively. 

Last, but not least, \textbf{RI4} is linked only to \textbf{RI1} through \textbf{TS12} and can be dissociated into the enol CH$_2$CHOH + H or into the epoxide CH$_2$OCH$_2$ + H. These steps exhibit 110 kJ/mol (\textbf{TS13}) and 250 kJ/mol (\textbf{TS14}) barriers, respectively.

To summarize, the possible products are, in order of stability: formaldehyde H$_2$CO + CH$_3$, acetaldehyde CH$_3$CHO + H, CH$_2$CH + H$_2$O, ethene CH$_2$CH$_2$ + OH, the enol CH$_2$CHOCH + H and the epoxide CH$_2$OCH$_2$ + H.

It is noteworthy that the first addition is barrierless and that the energies of all the involved intermediates and TSs are below that of the reactants, which makes this path viable in ISM, and that acetaldehyde CH$_3$CHO is among the possible products. A kinetics study will therefore be needed to figure out the amount actually formed via this reaction.

\textit{CH$_3$OH + CH.} Fig. \ref{fig:path-CH3OH+CH} shows the path representing the CH + CH$_3$OH reaction, only focusing on the intermediates and TSs that involves energies below that of the reactants, and therefore viable in ISM. Other addition or insertion first steps were considered but were too high energetically and therefore not pictured here.  The exhibited energies are again the CCSD(T)/aug-cc-pVTZ reevaluated electronic energies corrected with the B2PLYP-D3/aug-cc-pVTZ zero-point energies (ZPE). 

It is notable that this path is very similar to the previous one, due to the identical intermediates involved. The main differences on this new path are the higher energy of the reactants that shifts the path downwards and the existence of a Van der Waals complex \textbf{VdW} before the two TSs \textbf{TS-insCH} and \textbf{TS-insOH}. These TSs represent the insertion of CH inside of the C-H bond and the O-H bond of methanol and exhibit 42 and 33 kJ/mol barriers, respectively. The first intermediates here are therefore \textbf{RI2} and \textbf{RI4} and not \textbf{RI1} anymore. All these factors will play a role on the kinetics of the system, which is again needed to verify the efficiency of this path in forming acetaldehyde. All the transition states involved in both channels are depicted in Fig. \ref{fig:TSs} in Appendix. 

\subsection{Kinetics results}
\label{subsec:kinetics-results} 

\begin{figure*}
	\includegraphics[scale=0.65]{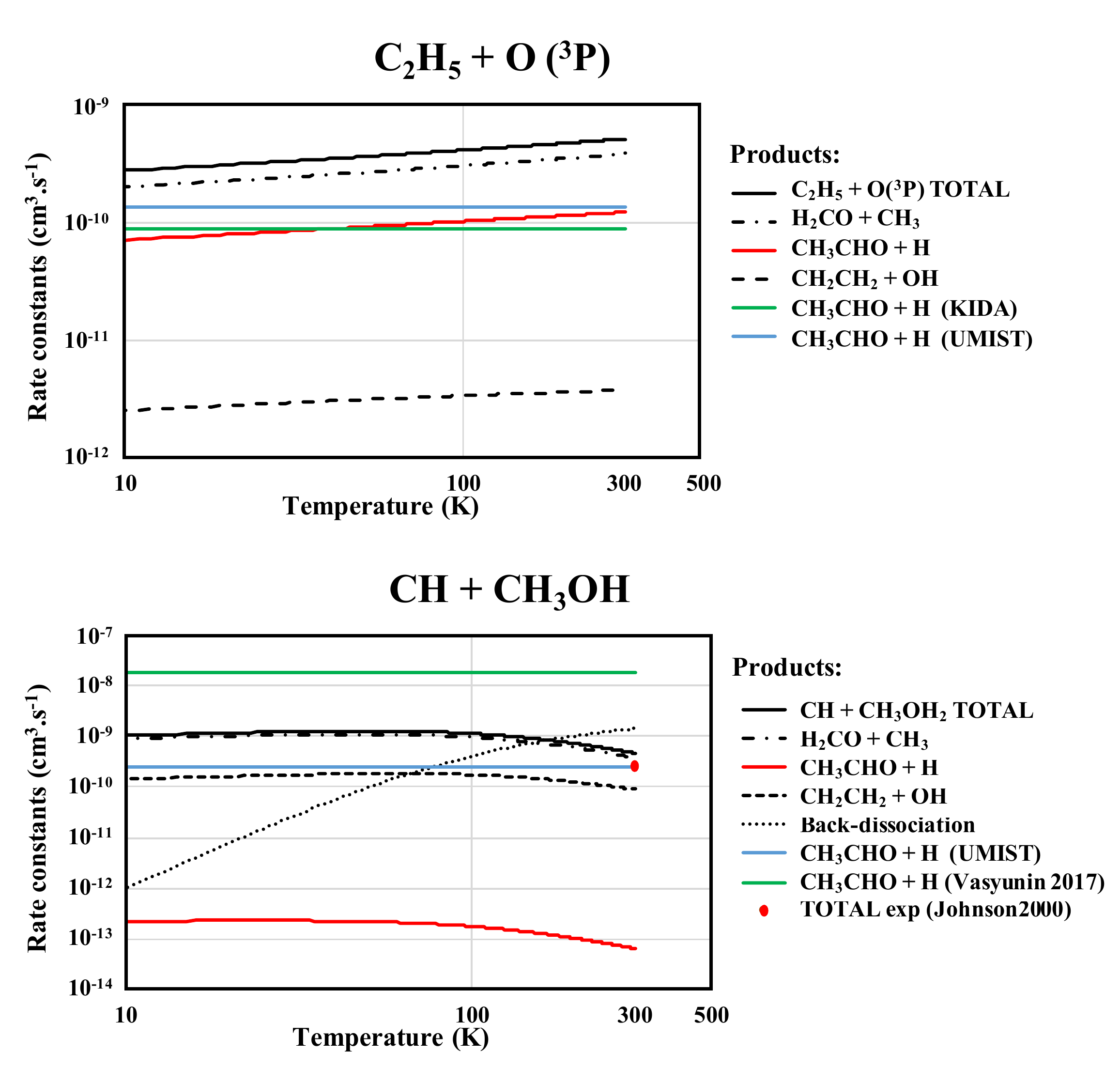}
    \caption{Rate constants as a function of temperature for the formation of the major products of the C$_2$H$_5$ + O($^3$P) and CH$_3$OH + CH reactions, respectively.}
    \label{fig:rates}
\end{figure*}

The RRKM method was used to evaluate the rate constants of the formation of the major products of each reaction, and, more particularly, acetaldehyde. The results are shown in Fig. \ref{fig:rates}.

\textit{C$_2$H$_5$ + O($^3$P).} A factor 2/3 was applied to the rate constants obtained for this reaction, based on the work by \cite{harding2005}, which stated that out of three states, only two are reactive for this system. One can see on Fig. \ref{fig:rates} that the major products of this reaction should be H$_2$CO + CH$_3$, directly followed by CH$_3$CHO + H, at any temperature. This can be explained by the facts that only one step is required to reach them from \textbf{RI1} and that the TSs that need to be over-passed are quite low in energy. It is an encouraging result, as acetaldehyde is the compound of interest here. The back-dissociation into the reactant is negligible due to the huge stabilization of the first intermediate (\textbf{RI1}, by 366.2 kJ/mol). 

\textit{CH + CH$_3$OH.} As far as the second reaction is concerned, H$_2$CO + CH$_3$ and CH$_2$CH$_2$ are supposed to be the major products at low temperatures, but when the temperature is increasing (after \textit{ca.} 170 K), back-dissociation becomes prevailing due to the existence of the Van der Waals complex that can, quite easily, dissociate into the reactants. It is noticeable that, unfortunately, acetaldehyde is formed in a negligible amount, as it requires several steps including quite high transition states in the reaction to be reached.

In order to be used by astrochemical models, we fitted the computed rate constants between 7 and 300 K with the function $k(T)=\alpha({\frac{T}{300K}})^\beta$, leaving $\alpha$ and $\beta$ as free parameters.
Please note that, in order to obtain a better fitting, we split the temperature in two ranges, above and below 95 K.
The obtained values of $\alpha$ and $\beta$ are reported in Table \ref{tab:alpha-beta} for the major formation products of both C$_2$H$_5$ + O($^3$P) and CH$_3$OH + CH reactions. 

\begin{table*}
    \begin{center}
	\caption{Summary of the computed reaction rate constants as a function of the temperature of the major products of the C$_2$H$_5$ + O($^3$P) and CH$_3$OH + CH reactions studied in this work. 
	The rate constants were fitted with the function $k(T)=\alpha({\frac{T}{300K}})^\beta$, used in astrochemical models. 
	Columns 2 and 3 report $\alpha$ and $\beta$ in two temperature ranges (column 4) , 7-95 K and 95-300 K, chosen for a better fit than that obtained with only one range. 
	Columns 5 and 6 list the reaction rate constants (in cm$^3$ s$^{-1}$) computed at 10 K and 100 K, namely the approximate temperatures of cold molecular clouds and a hot cores/corinos, respectively. 
	Finally, for comparison, columns 7 and 8 quote the values given in the KIDA \citep{Talbi2011,Wakelam2012} and UDfA \citep{Woodall2007,mcelroy2013} databases.}
	\label{tab:alpha-beta}
	\begin{tabular}{lccccc|cccc} 
		\hline
 & \multicolumn{5}{c}{This study} & \multicolumn{2}{c}{KIDA} & \multicolumn{2}{c}{UDfA} \\
Reaction & $\alpha$ [cm$^3$ s$^{-1}$] & $\beta$ & T [K] & k$_{10 K}$ [cm$^3$ s$^{-1}$] & k$_{100 K}$ [cm$^3$ s$^{-1}$] & $\alpha$ [cm$^3$ s$^{-1}$]& $\beta$ & $\alpha$ [cm$^3$ s$^{-1}$] & $\beta$ \\
		\hline
C$_2$H$_5$ + O($^3$P) $\rightarrow$ CH$_3$CHO + H & $1.21\times10^{-10}$ & 0.16 & 7-300 & $0.71\times10^{-10}$ & $1.02\times10^{-10}$ & $8.80\times10^{-11}$ & 0 & $1.33\times10^{-10}$ & 0  \\
C$_2$H$_5$ + O($^3$P) $\rightarrow$ H$_2$CO + CH$_3$ & $3.65\times10^{-10}$ & 0.18 & 7-95 & $1.99\times10^{-10}$ & $3.03\times10^{-10}$ & $6.60\times10^{-11}$ & 0 & $2.67\times10^{-11}$ & 0 \\
    & $3.82\times10^{-10}$ & 0.21 & 95-300 \\
C$_2$H$_5$ + O($^3$P) $\rightarrow$ CH$_2$CH$_2$ + OH & $3.87\times10^{-12}$ & 0.13 & 7-95 & $2.51\times10^{-12}$ & $3.35\times10^{-12}$ & $4.40\times10^{-11}$ & 0 & - & - \\
    & $3.75\times10^{-12}$ & 0.10 & 95-300 \\
    \hline
    \hline
CH$_3$OH + CH $\rightarrow$ CH$_3$CHO + H & $1.84\times10^{-13}$ & -0.07 & 7-95 & $2.21\times10^{-13}$ & $1.75\times10^{-13}$ & - & - & $2.49\times10^{-10}$ & -1.93 \\
    & $6.74\times10^{-14}$ & -0.95 & 95-300 \\
CH$_3$OH + CH $\rightarrow$ H$_2$CO + CH$_3$ & $1.16\times10^{-9}$ & 0.06 & 7-95 & $9.02\times10^{-10}$ & $9.62\times10^{-10}$ & - & - & - & - \\
    & $4.00\times10^{-10}$ & -0.88 & 95-300 \\
CH$_3$OH + CH $\rightarrow$ CH$_2$CH$_2$ + OH & $2.14\times10^{-10}$ & 0.10 & 7-95 & $1.44\times10^{-10}$ & $1.75\times10^{-10}$ & - & - & - & - \\
    & $9.40\times10^{-11}$ & -0.63 & 95-300 \\
		\hline
	\end{tabular}
	\end{center}
\end{table*}

\section{Discussion}
\label{sec:discussion}

\subsection{A new network for the formation of acetaldehyde}\label{subsec:chemical-network}

In the literature, four gas-phase formation routes of acetaldehyde have been invoked (see \S ~\ref{sec:reviews}): path 1, following the recombination of the protonated acetaldehyde (CH$_3$CHOH$^+$); paths 2 and 3, via reactions of ethyl radical (C$_2$H$_5$) and methanol (CH$_3$OH) with O and CH, respectively; path 4, starting from ethanol (see the summary in Fig. \ref{fig:scheme-react}).
In this study and in two previous works \citep{vazart2019,Skouteris2018}, we studied the four reactions via theoretical computations of the electronic energy (\S ~\ref{subsec:paths}) and kinetics (\S ~\ref{subsec:kinetics-results}).
Table ~\ref{tab:alpha-beta} summarizes the results of the new computations of the present work. 
We report the $\alpha$ and $\beta$ parameters, from which to compute the rate constants as a function of the temperature, and Fig. ~\ref{fig:rates} plots them. 
With this new study, therefore, we are now in a position to assess which path efficiently can form acetaldehyde and, perhaps, is responsible for its presence in the ISM.

\medskip \noindent
{\it Path 1:} CH$_3$OCH$_3$ + H$^+$ \& CH$_3$CHOH$^+$ + e$^-$\\
The study of this ionic route was reported in \cite{vazart2019}, where we showed that the reaction supposed to lead to protonated acetaldehyde actually does not form it. 
In fact, the reaction leads to the formation of CH$_2$OH$^+$ + CH$_4$ rather than CH$_3$CHOH$^+$ + H$_2$, as previously quoted in astrochemical databases.
Consequently, since there are not known routes that efficiently form protonated acetaldehyde, the ionic formation route of acetaldehyde is invalid. 
Note that \cite{Skouteris2018} had already removed this route in their chemical network, suspecting an improbable rearrangement of the atoms for the reaction to occur. 

\medskip \noindent
{\it Path 2:} C$_2$H$_5$ + O\\
This study claims that, although acetaldehyde is not the major product (formaldehyde is), acetaldehyde is synthesised at a few times $10^{-10}$ cm$^3$ s$^{-1}$, a rate constant almost unchanged between 7 and 300 K. 
The reaction forms about three times more formaldehyde and thirty times less ethylene (CH$_2$CH$_2$).
Indeed, the transition state leading to acetaldehyde is slightly higher in energy than the one leading to formaldehyde (by 19 kJ/mol, cf. Fig. \ref{fig:path-C2H5+O}), which leads to the rate constant for the formation of the latter slightly faster, as seen in Fig. \ref{fig:rates} and to a branching ratio (BR) of \textit{ca.} 30\%.\\ 
Commonly, a 40-50\% BR is given in the literature regarding the formation of acetaldehyde by this reaction \citep{slagle1988,hoyermann1999,hack2002,harding2005}, and this can be explained using a temperature argument. Indeed, in \cite{hoyermann1999}, comparable quantum chemistry computations were performed regarding a few steps of the reaction and are in agreement with the present study (the energy difference between the transition states leading to acetaldehyde and formaldehyde being of 21 kJ/mol). They also performed a kinetics study, starting from room temperature, that shows that at their lowest temperatures, formaldehyde is the major product of the reaction. But when the temperature increases, both formaldehyde and acetaldehyde tend to be formed at the same rate, which can explain why the 40-50\% BR is ordinarily reported in the literature.

\medskip \noindent
{\it Path 3:} CH$_3$OH + CH\\ 
This reaction forms mainly formaldehyde and ethylene and only a negligible fraction ($\sim 2\times10^{-4}$) of acetaldehyde. 
Indeed, the first intermediate requires several rearrangements in the reaction path to form acetaldehyde, while formaldehyde and ethylene can be formed after a direct dissociation of this intermediate. 
Moreover, the presence of a Van der Waals complex before the formation of the first intermediates leads to the significant role of back-dissociation at temperatures higher than about 170 K and, therefore, to a decrease of the formation rate constants after this temperature.
As a result, only 1/5000 times one acetaldehyde molecule is formed, with a rate constant ($\sim 2\times10^{-13}$ cm$^3$ s$^{-1}$), which is about 650 times lower than the one from reaction (2). 
In other words, this route of formation of acetaldehyde is very likely negligible, except in environments where ethyl radical or atomic oxygen are more than 650 times less abundant than methanol and CH, which is an unlikely situation (see the discussion in \S ~\ref{subsec:astro}).

\medskip \noindent
{\it Path 4:} CH$_3$CH$_2$OH + OH \& CH$_3$CHOH + OH\\ 
This path that leads (also) to the formation of acetaldehyde was studied via theoretical computations by \cite{Skouteris2018}. 
The goal of that study was to show a gas-phase route to glycolaldehyde, but acetaldehyde is a by-product of what was called "the genealogical ethanol tree", as from the reaction of ethanol with OH and Cl, other three iCOMs can be formed (formic acid, acetic acid and acetaldehyde).
The overall rate constant of acetaldehyde formation from the ethanol tree is large enough to make it a potential source of interstellar acetaldehyde.

\begin{figure*}
	\includegraphics[scale=0.47]{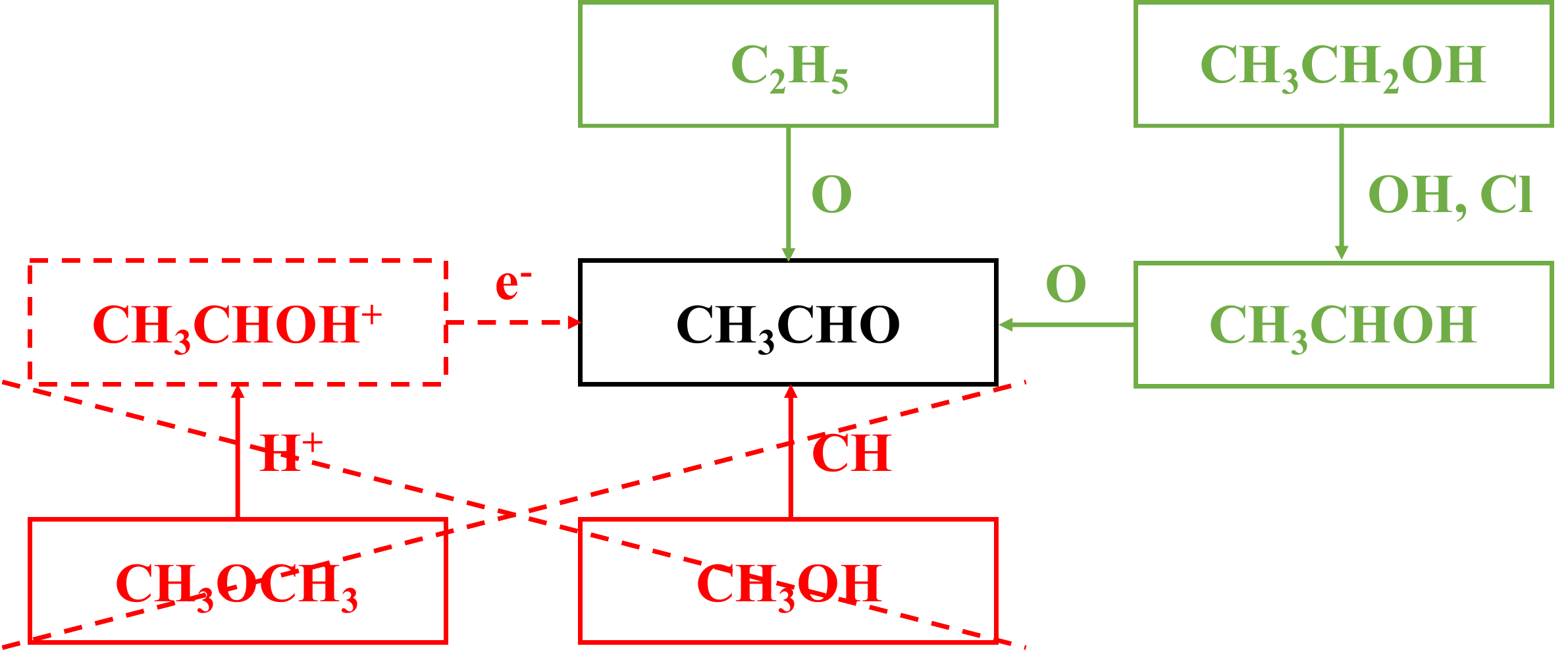}
    \caption{Scheme of the gas-phase formation of acetaldehyde according to the reactions proposed in the literature before the present study (\S ~\ref{sec:reviews}). 
    The color code indicates whether the reaction is validated (green) or disproved (red) by our new and old computations (see text).}
    \label{fig:summary}
\end{figure*}

\medskip
In conclusion, as summed up in Fig. \ref{fig:summary}, among the four most potentially important gas-phase formation routes of acetaldehyde invoked in the literature, only two will be efficient in the ISM conditions, based on our computations: the paths 2 and 4 described in Section ~\ref{sec:reviews}.
We encourage the astrochemical modellers to use only these two reaction paths for the acetaldehyde formation in the gas-phase. 

\subsection{Comparison of our new computation results with previous studies and astrochemical databases}\label{subsec:comp-previous}

In this  section, we review how our new computations compare with experimental and previous theoretical values and with the values reported in the KIDA and UDfA databases, largely used in astrochemical models.

\medskip \noindent
{\it Reaction 2:} C$_2$H$_5$ + O\\
To the best of our knowledge, no experimental data are available in the literature. The KIDA and UDfA databases report constant values equal to $8.8\times10^{-11}$ cm$^3$ s$^{-1}$ and 1$.33\times10^{-10}$ cm$^3$ s$^{-1}$, respectively. These values are taken from \cite{Baulch2005} and \cite{Hebrard2009} in KIDA, and from NIST in UDfA.
These values compare extremely well with those computed in this work ($1.21\times10^{-10}$ cm$^3$ s$^{-1}$), especially the ones reported in UDfA.
Having said that, both databases present acetaldehyde as the major product of the reaction, which is not correct.  

\medskip \noindent
{\it Reaction 3:} CH$_3$OH + CH\\
The only published experiment on this reaction is the one by \cite{Johnson2000}, who studied the global rate constant between 298 and 753 K and at a 100-600 Torr pressure, but they could not distinguish the different products of the reaction because their technique does not provide it.
In the temperature range of their study, Johnson et al. found a steep dependence on the temperature, -1.93.
Our new computations compare relatively well, within a factor two, with the Johnson et al. global rate constant at 298 K, $4.65\times10^{-10}$ against the measured $(2.5\pm0.1)\times10^{-10}$ cm$^3$ s$^{-1}$.
We did not carry out computations in the range studied by \cite{Johnson2000} but the shape of our curve suggests a decrease of the rate constants when the temperature increases, due to the prevalence of back-dissociation after 170 K, which coincides with their results.

Reaction 3 is reported in the UDfA database to form acetaldehyde with $\alpha$ equal to $2.49\times10^{-10}$ cm$^3$ s$^{-1}$ and $\beta$ equal to -1.93, (erroneously) based on the work by \cite{Johnson2000}. 
However, as said, the Johnson et al. value refers to the global rate constant of the reaction, not to the formation of acetaldehyde, and the temperature dependence is in a totally different range. 
On the contrary, we found more than 1000 times lower value of $\alpha$, $1.84\times10^{-13}$ cm$^3$ s$^{-1}$, compared to the value reported in UDfA, and an almost null dependence on the temperature at $\leq 95$ K.
Note that the KIDA database does report reaction 3 but assumes that it leads to the formation of CH$_3$ + H$_2$CO only, with a constant rate constant of $2.5\times10^{-10}$ cm$^3$ s$^{-1}$, which is indeed what we find.
Finally, \cite{Vasyunin2017} proposed this reaction based on the experimental study by \cite{Johnson2000} and assumed that 10\% of the reaction leads to acetaldehyde, which is wrong, and the remaining 90\% to formaldehyde.
They also kept the -1.93 temperature dependence so that they assumed an acetaldehyde formation rate constant of $1.8\times10^{-8}$ cm$^3$ s$^{-1}$ at 10 K, about five orders of magnitude larger than the values computed by us (and unreasonably high for a neutral-neutral reaction at those temperatures).

\begin{table*}
    \begin{center}
	\caption{CH$_3$CHO abundances measured towards hot cores, prestellar cores, hot corinos, and protostellar shocks} 
	\label{tab:detections}
\begin{tabular}{lccccc} 
	\hline
 Source & T (K) & N (cm$^{-2}$) & Abundance/H$_2$ & N$_{H_2}$ (cm$^{-2}$) & Reference\\
	\hline
 \multicolumn{6}{c}{\it Hot cores}\\
 	\hline
 	\hline
G34.43+00.24 MM3 & 110 & 3.4 $\times$ 10$^{14}$ & - & - & [1]\\
G328.2551-0.5321 & 90-110 & 6.8 $\times$ 10$^{15}$& 3.6 $\times$ 10$^{-9}$ &  1.9 $\times$ 10$^{24}$ &[2]\\
Sgr B2 (N2) & 150 & 4.3 $\times$ 10$^{17}$ & 2.7 $\times$ 10$^{-7}$ &  1.6 $\times$ 10$^{24}$ &  [3]\\
Sgr B2 (N3) & 145 & 8.5 $\times$ 10$^{16}$ & 9.4 $\times$ 10$^{-8}$ &  0.9 $\times$ 10$^{24}$ &  [3]\\
Sgr B2 (N4) & 145 & 9.0 $\times$ 10$^{16}$ & 3.5 $\times$ 10$^{-8}$ &  2.6 $\times$ 10$^{24}$ &  [3]\\
Sgr B2 (N5) & 145 & 2.5 $\times$ 10$^{16}$ & 2.8 $\times$ 10$^{-8}$ &  0.9 $\times$ 10$^{24}$ &  [3]\\

\hline
 \multicolumn{6}{c}{\it Prestellar cores}\\
 \hline
 	\hline
L1544  & 17 &  5 $\times$ 10$^{11}$ & 1 $\times$10$^{-10}$&  5 $\times$ 10$^{21}$& [4]\\
L1544 continuum peak & 5 & 1.2$\times$10$^{12}$ & 2.2$\times$10$^{-11}$ & 5.4$\times$10$^{22}$ & [5]\\
L1544 methanol peak & 7.8 & 3.2$\times$10$^{12}$ & 2.1$\times$10$^{-11}$ & 1.5$\times$10$^{22}$ & [5]\\
Barnard 5 & 5 & 5.2 $\times$ 10$^{12}$ & 1.6 $\times$ 10$^{-9}$& 3.3 $\times$ 10$^{21}$& [6]\\
Taurus cores & 3-5  & 0.7-5.8 $\times$ 10$^{12}$ & - & - & [7]\\

\hline
 \multicolumn{6}{c}{\it Hot corinos}\\
 \hline
 	\hline
Barnard 1b-S (hot) & 200 & 8 $\times$ 10$^{14}$ & 5.7 $\times$ 10$^{-11}$& 1.4 $\times$ 10$^{25}$& [8]\\
Barnard 1b-S (cold) & 60 & 1.6 $\times$ 10$^{14}$ & 1.4 $\times$ 10$^{-11}$&1.1 $\times$ 10$^{25}$& [8]\\
HH212 & 78 & 8$\times$10$^{15}$ & 8$\times$10$^{-9}$& 10$^{24}$& [9]\\
B335 & 100 &14 $\times$ 10$^{14}$ & 24$\times$10$^{-10}$& 6 $\times$ 10$^{23}$& [10]\\
IRAS16293-2422 & 70 &  1 $\times$ 10$^{15}$ & 0.3 $\times$10$^{-8}$&  3 $\times$ 10$^{23}$ &  [11]\\
IRAS16293-2422 B  & 140 &  3.5 $\times$ 10$^{15}$ & - & - & [12]\\
NGC1333-IRAS4A2 & 100-200 & (1.0 - 1.9) $\times$ 10$^{16}$ & (1.1 - 7.4) $\times$ 10$^{-9}$& (1.9 - 2.7) $\times$ 10$^{24}$& [13]\\
L483 & 100-300 & 8 $\times$ 10$^{16}$ & - & - & [14]\\
SVS13-A & 35 & 12 $\times$ 10$^{15}$ & 4 $\times$ 10$^{-9}$& 3 $\times$ 10$^{24}$& [15]\\
\hline
 \multicolumn{6}{c}{\it Protostellar shocks}\\
 \hline
 	\hline
L1157-B1b & 90 & 5 $\times$ 10$^{15}$ & 2.5 $\times$ 10$^{-6}$& 2 $\times$ 10$^{21}$& [16]\\
NGC1333-IRAS4A outflow & 9-30 & 0.2-1.3 $\times$ 10$^{14}$ & - & - & [17]\\ 
		\hline
	\end{tabular}
	\end{center}
[1] \citet{Sakai2018}; 
[2] \citet{Csengeri2019};
[3] \citet{Bonfand2019};
[4] \citet{Vastel2014}; 
[5] \citet{Jimenez2016}; 
[6] \citet{Taquet2017};
[7] \citet{Scibelli2020};
[8] \citet{Marcelino2018};
[9] \citet{Lee2019,Codella2019};
[10] \citet{Imai2016}; 
[11] \citet{Jaber2014};
[12] \citet{Manigand2020, Jorgensen2018, Jorgensen2016};
[13] \citet{Lopez2017};
[14] \citet{Jacobsen2019};
[15] \citet{Bianchi2019};
[16] \citet{Codella2020}; 
[17] \citet{DeSimone2020}\\
\end{table*}

\subsection{Astronomical observations}\label{subsec:astro}

Acetaldehyde was one of the earliest molecules to be detected in space. 
After the first detection towards the galactic center in 1973 \citep{Gottlieb1973}, acetaldehyde has been detected in many star formation environments and in a large range of interstellar conditions, as summarised in Tab. ~\ref{tab:detections}. 
This indicates that CH$_3$CHO is efficiently formed at all gas temperatures, from  about 10 K in prestellar cores up to temperatures larger than 200 K in hot cores. 
However, the measured CH$_3$CHO abundances vary by five orders of magnitude, from $\sim$ 10$^{-11}$ in cold environments (e.g. prestellar cores) to up to $\sim10^{-6}$ in warm ones (e.g. hot cores/corinos and protostellar shocks).
This may or may not indicate that a different chemical route is responsible for the formation of acetaldehyde in different environments.
In the following, in order to understand better this point, we will review the possible formation routes in cold and warm environments, respectively, in comparison with astronomical observations.

\subsubsection{Cold environments}\label{subsubsec:cold-env}
Despite the low temperatures (< 10 K), acetaldehyde is commonly observed in starless and prestellar cores \citep[e.g.]{Bacmann2012, Cernicharo2012, Vastel2014, Jimenez2016}. Recently, \citet{Scibelli2020} performed a survey of starless and prestellar cores in the Taurus molecular cloud and detected CH$_3$CHO in about 70 \% of the sample sources.

Two major paths have been invoked to explain gaseous acetaldehyde in these cold environments.
The first one relies on three steps \citep[e.g.][]{Vasyunin2013,Vastel2014,Jimenez2016}: (1) the formation of the ethyl radical on the grain surfaces, by hydrogenation of small hydrocarbons formed in the gas-phase and frozen on the grain surfaces at later stages; (2) the injection of ethyl radical in the gas phase via (perhaps) chemical desorption (see below); (3) gas-phase formation of acetaldehyde via the reaction of ethyl radical  with atomic oxygen (our reaction 2). 
In this scheme, the last step is now validated.

The second path, which was proposed later in the literature \citep{Vasyunin2017,Scibelli2020}, invokes: (1) the formation of methanol by CO hydrogenation on the grain surfaces; (2) the injection of methanol in the gas phase via (perhaps) chemical desorption; (3) gas-phase formation of acetaldehyde via the reaction of methanol with CH (our reaction 3). 
As discussed in \S ~\ref{subsec:chemical-network}, the last step of this path occurs at a rate constant which is about five orders of magnitude lower than that used by astrochemical models \citep{Vasyunin2017} and, consequently, ineffective in reproducing the observed abundance in cold environments. 
This path would be competitive with respect to the first one only if the bottleneck reactant of reaction (3), methanol, is about 650 times more abundant of that of reaction (2), ethyl radical. 
Since no radio to millimeter wavelengths are available to identify ethyl radical in space, we cannot compare measured abundances of ethyl radical and methanol.
However, if one then relies on model predictions, the latter do not support  650 times more methanol then ethyl radical.
In addition, would gaseous methanol be 650 times more abundant than ethyl radical, reaction (3) would produce a formaldehyde abundance two orders of magnitude larger than that observed.
Therefore, the chemical path proposed by \cite{Vasyunin2017} is completely excluded by our new computations and we recommend to drop it from astrochemical models.

In conclusion, based on our new computations, the first path, which involves reaction (2), is the only viable one of the two invoked in the literature.
The weak ring of this formation chain remains the second step, namely the chemical desorption.
In this process, it is assumed that part of the energy released by the hydrogenation on the grain surfaces is acquired by the newly formed species (in this case C$_2$H$_5$) which can then break its bonds with the surface and be liberated into the gas phases \cite[see. e.g.,][]{duley1993}. 
The fraction of released species, however, strongly depends on the species and substrate \citep{minissale2016,Oba2018} and could be null for relatively low reaction energies and strongly bound species \citep{Pantaleone2020}. 
In the specific case of C$_2$H$_5$, there are no experiments or theoretical computations providing estimates or even just constraints. 
We only can say that the comparison of astrochemical model predictions with observations suggest that a relatively small C$_2$H$_5$ abundance, of about $5\times10^{-9}$ with respect to molecular hydrogen, will be enough to reproduce the observations \citep{Vastel2014}.

\subsubsection{Warm environments}\label{subsubsec:warm-env}
Abundant acetaldehyde is routinely observed in high-mass hot cores \citep[e.g.][]{Blake1987, Sakai2018, Csengeri2019, Bonfand2019}, low-mass hot corinos \citep[e.g.][]{cazaux2003, Jorgensen2018, Bianchi2019, Lee2019, Codella2018, Codella2019}, and protostellar molecular shocks \citep{lefloch2017,codella2017,Codella2020, DeSimone2020}.

In this case, three paths have been invoked in the literature to explain the observations.
In the first one, everything happens on the grain surfaces in four steps \citep[e.g.][]{Garrod2006,oberg2016}:
(1) the freeze-out of species and their hydrogenation; (2) the formation of frozen radicals, in this specific case CH$_3$ and HCO, by the UV illumination of the frozen hydrogenated species; (3) when the dust warms up because of the presence of the protostar, the CH$_3$ and HCO radicals diffuse inside the grain ices, meets and combine into acetaldehyde; (4) frozen acetaldehyde is injected into the gas-phase when the dust temperature reaches the sublimation temperature of the ices or the shock sputters the ice content.
Although there is no doubt about the last step viability, namely the thermal desorption and shock sputtering, the formation of acetaldehyde from the HCO + CH$_3$ combination on the grain icy mantles (step 3) is source of debate.
Theoretical calculations show that the combination of CH$_3$ and HCO on the ice surfaces does not necessarily lead to acetaldehyde because the reaction possesses a non-negligible activation energy (up to 6.8 kcal/mol depending on the position on the ice surface) and it is in competition with the formation of CH$_4$ + CO \citep{Enrique-Romero2019,Enrique-Romero2020}.
In addition, comparison of high spatial resolution acetaldehyde observations towards protostellar shocks privileges the gas-phase formation of acetaldehyde via the C$_2$H$_5$ + O reaction in these environments \citep{DeSimone2020,Codella2020}.

\begin{figure*}
\begin{center}
	\includegraphics[scale=0.8]{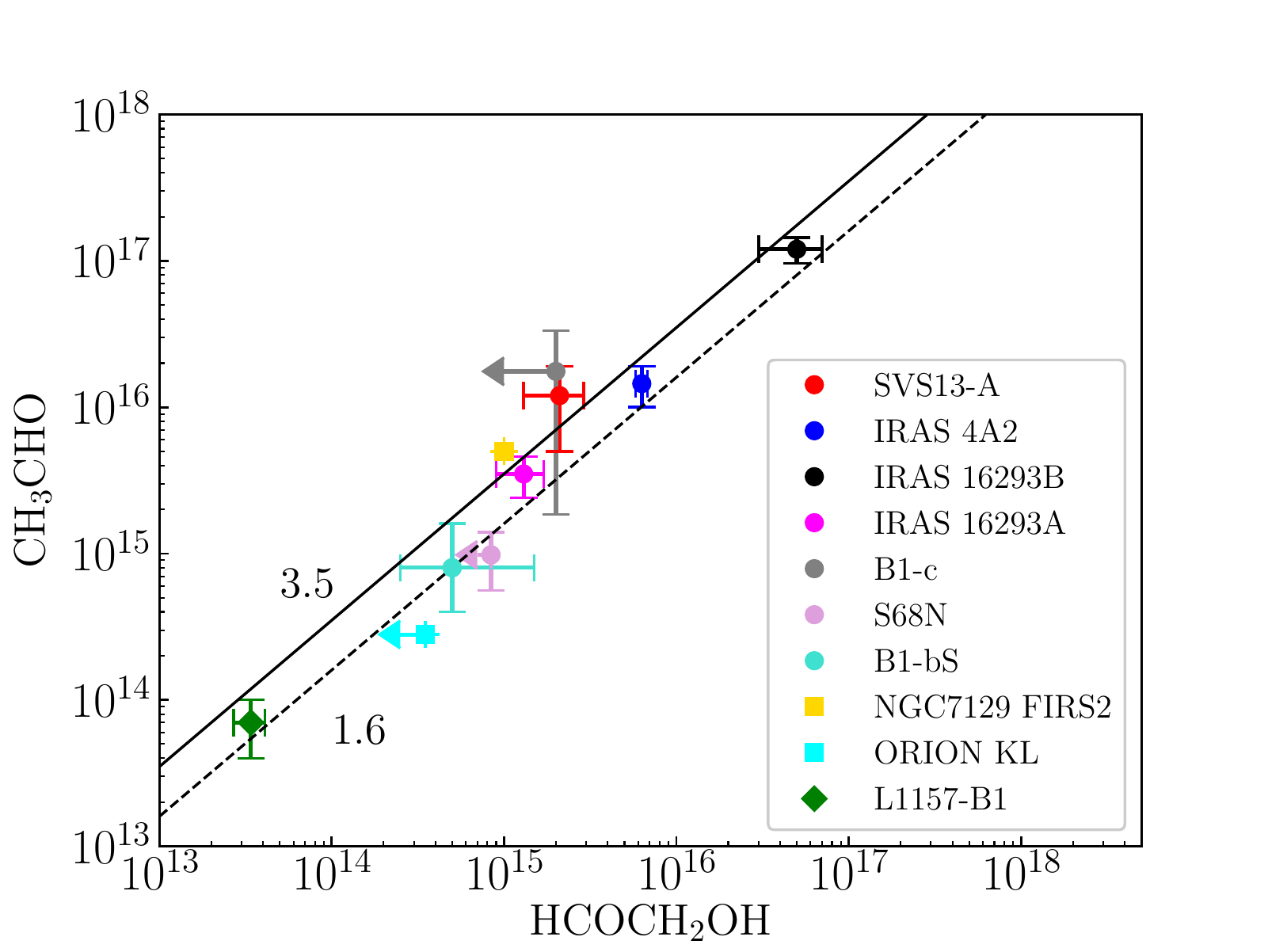}
    \caption{Column densities of glycolaldehyde (HCOCH$_2$OH) and acetaldehyde (CH$_3$CHO) measured in several astronomical sources, marked in the inset.
    The two lines show the expected ratio if acetaldehyde and glycolaldehyde are both synthesised in the gas-phase the via the ethanol tree by \citet{Skouteris2018}. 
    The solid and dashed lines refer to the uncertain value of the first step of the path and correspond to an acetaldehyde over glycolaldehyde abundance ratio of 3.5 and 1.6, respectively.
    The references for the plot are: \citet{Bianchi2019, Desimone2017, Lopez2017,Jorgensen2016,Jorgensen2018, Marcelino2018, Manigand2020, vanGelder2020,Fuente2014} and references therein; \citet{lefloch2017, Codella2020}.}
    \label{fig:obs-aceta-glyco}
    \end{center}
\end{figure*}

The second invoked path is the same as the first one in cold environments, described above, namely it involves the reaction (2) (C$_2$H$_5$ + O) \citep[e.g.][]{charnley2004}.
The difference in the two schemes is that the chemical desorption is replaced by the thermal desorption and the shock sputtering, both proved processes.
We have seen that our computations validate reaction (2), so that the acetaldehyde gas-phase formation path is fully viable.
The most uncertain point is the abundance of ethyl radical.
As said earlier, since its frequencies and spectroscopic data are not known yet, there is no way to verify that the route of acetaldehyde formation from ethyl radical is the true one.

\subsubsection{A new actor on the scene: the ethanol tree}\label{subsubsec:eth-tree}

The path 4, the ethanol tree, one branch of which is acetaldehyde, was introduced and studied by \cite{Skouteris2018}.
They showed that about 6\% of this reaction path ends up into acetaldehyde.
Mostly interesting, this is between 1.6 and 3.5 times the branching ratio of the formation of glycolaldehyde (HCOCH$_2$OH) (please note that the uncertainty comes from the uncertainty in the first step of the path: see \citet{Skouteris2018}), which was the focus of the study.
It is worth reminding that the ethanol tree is the only known path to the formation of glycolaldehyde in the gas-phase and that the comparison of model predictions including this path and the observed values are in agreement, so far \citep{Skouteris2018}.
Successive works have also showed a tight correlation between glycolaldehyde and ethanol, extending that found by \citet{Skouteris2018} toward the low end by more than one order of magnitude \citep{Li2019,Xue2019} and, consequently, strengthening the validity of this scheme in the glycolaldehyde production.
Although no author, not even the ethanol tree proposers, ever thought of it as an important source of acetaldehyde and, therefore, no specific modeling has been carried out so far, it is worth considering it here.

To this end, the easiest and most straightforward way to understand whether the ethanol tree is a competitive path for the formation of acetaldehyde is to compare the observed abundances of acetaldehyde and glycolaldehyde.
If the ethanol tree is the major source of acetaldehyde then the abundance ratio between the two species should be equal to the ratio of their respective path branching ratios, namely acetaldehyde should be within 1.6 and 3.5 times more abundant than glycolaldehyde.
Figure \ref{fig:obs-aceta-glyco} shows the measured column densities of glycolaldehyde and acetaldehyde towards several astronomical sources.
Please note that, since it is very difficult to derive the column density of H$_2$ and, therefore, reliable abundances, we plot the column densities, whose error bars are reported in the plot.
In the figure, we also show the theoretical ratio if both glycolaldehyde and acetaldehyde are formed via the ethanol tree.
The agreement between the predicted and the measured ratios is spectacular and strongly suggests that acetaldehyde is a daughter of ethanol.

The observations of Fig. \ref{fig:obs-aceta-glyco} refer to warm objects only as no glycolaldehyde has been detected in cold ones so far.
However, when we take into account that (i) the brightest glycolaldehyde line in the 70--150 GHz band, where cold objects are observed, is twice weaker than the one from acetaldehyde assuming the same column density and temperature for the two species and (ii) the column density of glycolaldehyde is 1.6-3.5 times smaller, the brightest glycolaldehyde lines would be between 3 and 7 times weaker than the acetaldehyde ones.
Therefore, the present non-detection upper limits to glycolaldehyde abundance in cold environments is compatible with the ethanol tree predictions so far.
For example, in L1544, one prestellar core where the full 3mm band was surveyed with IRAM-30m  high sensitivity observations \citep{Lefloch2018}, acetaldehyde was detected with a signal-to-noise ratio of 6.5 \citep{Vastel2014}, which explains the non-detection of glycolaldehyde.
\cite{Jimenez2016} observed another, brighter position towards L1544 and also their non-detection of glycolaldehyde is compatible with the ethanol tree predictions.

\subsubsection{Conclusive remarks}\label{subsubsec:concl}
In summary, our new theoretical computations show that, among the four previously proposed gas-phase reactions described in Section \ref{sec:reviews}, only two are viable, the 2 and 4, as summarised in Fig. \ref{fig:summary}.
The observed difference in the acetaldehyde abundance between the cold and warm environments (Tab. \ref{tab:detections}) can be easily be attributed to the difference in the parent's abundance, ethyl radical and/or ethanol, respectively.
In cold environments, ethyl radical and/or ethanol would be present in small quantities, because only a small fraction of the frozen species is injected into the gas phase by a non-thermal process, probably chemical desorption.
In warm environments, on the contrary, all ethyl radical and/or ethanol, probably previously formed and frozen on the grain surfaces, would be injected into the gas phase.

In order to make progresses and assess whether and when the two paths are important in the acetaldehyde formation, the abundances of the parents should also be measured.
As said in \S ~\ref{subsubsec:cold-env}, this is presently impossible for the ethyl radical, the parent in path 2, since its rotational transition frequencies are unavailable.
About path 4, the ethanol tree, we found in \S ~\ref{subsubsec:eth-tree} that the observed ratio of acetaldehyde and glycolaldehyde in several warm sources compares spectacularly well with that predicted based on the branching ratios of these two species.
It remains to show that it holds also in cold environments.

Finally, it also holds the possibility that acetaldehyde is a grain-surface radical-radical product, with the caveats described in \S ~\ref{subsubsec:warm-env}.
Providing the final answer of what process dominates the formation of acetaldehyde and in what environment will require a careful modeling and comparison with an expanded set of observations.

\section{Conclusions}\label{sec:conclusions}

n this paper, we presented a critical review of the gas-phase formation routes of acetaldehyde invoked in the literature and reported in the two major astrochemical databases, KIDA and UDfA.
We found that four paths are potentially important summarised:

\begin{tabular}{ll}
     (1) & CH$_3$OCH$_3$ + H$^+$ $\rightarrow$ CH$_3$CHOH$^+$ + H$_2$\\
         & CH$_3$CHOH$^+$ + e$^-$ $\rightarrow$ CH$_3$CHO + H\\
     (2) & C$_2$H$_5$ + O($^3$P) $\rightarrow$ CH$_3$CHO + H\\
     (3) & CH$_3$OH + CH $\rightarrow$ CH$_3$CHO + H\\
     (4) & CH$_3$CH$_2$OH + OH $\rightarrow$ CH$_3$CHOH + H$_2$O\\
         & CH$_3$CHOH + O $\rightarrow$ CH$_3$CHO + OH\\
\end{tabular}

The first path, involving the electron recombination of protonated acetaldehyde, was previously studied and excluded by \cite{vazart2019} because the formation of protonated acetaldehyde actually does not occur.
The fourth scheme starts from ethanol and was theoretically studied and validated by \cite{Skouteris2018}.
It is called "ethanol tree" because glycolaldehyde and other iCOMs are also formed from ethanol.
The second and third paths were not validated by neither experimental or theoretical works in the ISM conditions, namely low temperature and pressure.

In this work, we investigated these two reaction paths via theoretical chemistry calculations, using a composite CCSD(T) and DFT method for the electronic structure and the RRKM scheme for the kinetics.
For both reactions, we provide the rate constants as a function of the temperature and the branching ratios, in the format used by astrochemical models, in Tab. \ref{tab:alpha-beta}.
Our new calculations validate the reaction (2) and the values quoted in KIDA and UDfA, with the one from UDfA closer to our computed values.
On the contrary, our computed rate constants of the reaction (3) are about five orders of magnitude lower than those reported in the UDfA database and used by some models.
We therefore rule out that this reaction has a role in the acetaldehyde formation, at any temperature.

In summary, we conclude that only two gas-phase reaction paths, the (2) and (4), are potentially important in the gas-phase acetaldehyde formation.
Finally, we reviewed the observations of acetaldehyde towards warm and cold objects and their formation routes, in the light of the above conclusions.
In warm sources, the measured abundance ratio between glycolaldehyde and acetaldehyde is exactly that predicted by the ethanol tree, namely the path (4).
On the other hand, \citet{Skouteris2018} showed that the glycolaldehyde abundance measured in warm objects is reproduced by astrochemical model predictions based on the ethanol tree.
We therefore conclude that, very likely, also acetaldehyde is mainly formed by it.
In order to definitively confirm this hypothesis and to verify its validity also in cold environments, the comparison between dedicated model predictions and an expanded observational data-set is necessary.

\section*{Acknowledgements}

This project has received funding from the European Research Council (ERC) under the European Union's Horizon 2020 research and innovation programme, for the Project “The Dawn of Organic Chemistry” (DOC), grant agreement No 741002.

\section*{Data availability}
The data underlying this article are available in the article and in its online supplementary material.




\bibliographystyle{mnras}
\bibliography{main} 



\clearpage 
\appendix

\section{Involved transition states}

\begin{figure*}
	\includegraphics[scale=0.5]{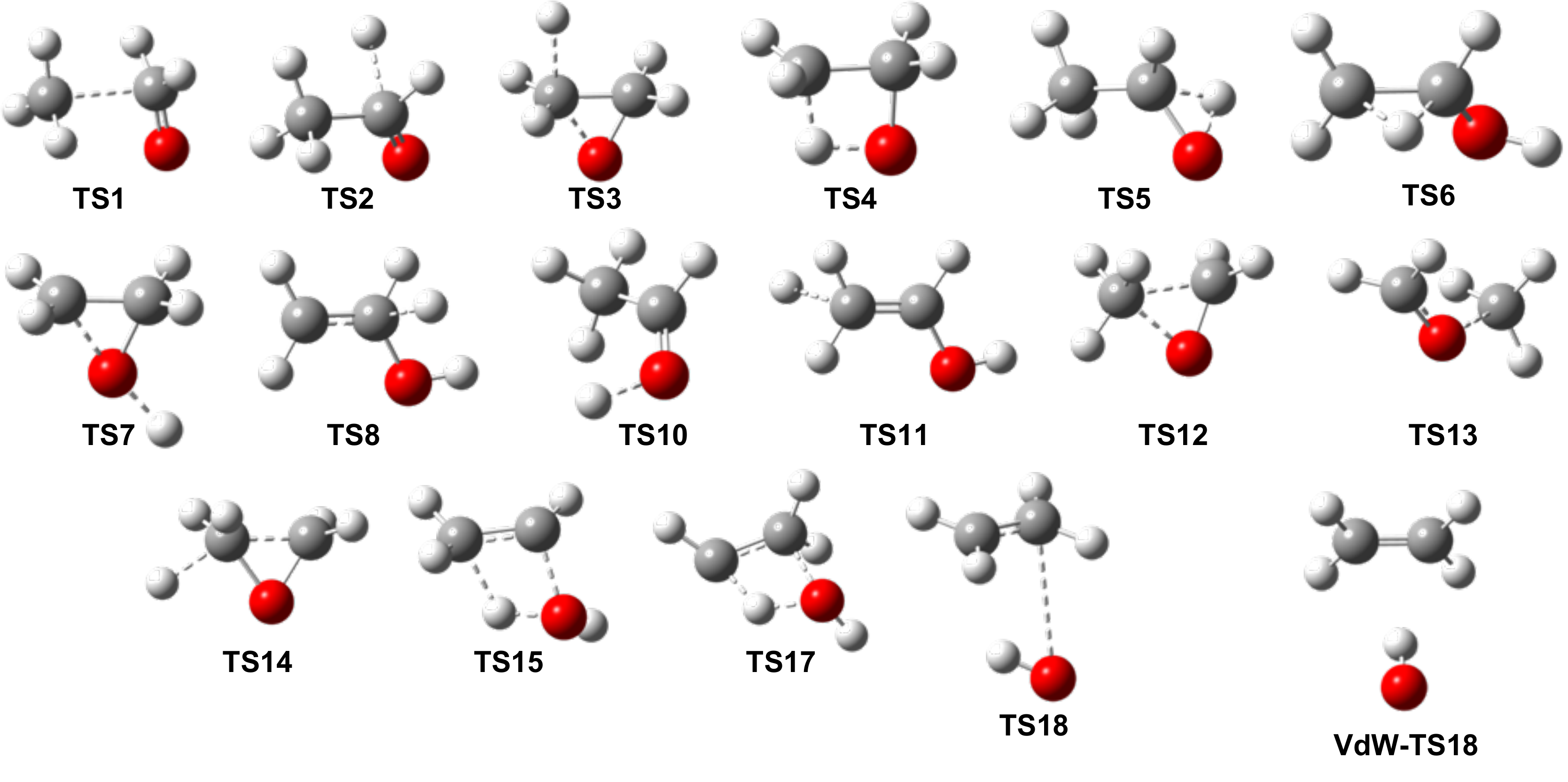}
    \caption{Transition states involved in both C$_2$H$_5$ + O($^3$P) and CH$_3$OH + CH channels.}
    \label{fig:TSs}
\end{figure*}

\section{Optimized geometries}

Level of theory : B2PLYP-D3/aug-cc-pVTZ

\textbf{C$_2$H$_5$}\\
 C \tabto{1cm}       0.009932   -0.692390    0.000000\\
 H \tabto{1cm}       0.504457   -1.095862    0.883276\\
 H \tabto{1cm}       0.504457   -1.095862   -0.883276\\
 H \tabto{1cm}      -1.009856   -1.098196    0.000000\\
 O \tabto{1cm}       0.009932    0.792120   -0.000000\\
 H \tabto{1cm}      -0.059118    1.345771   -0.922933\\
 H \tabto{1cm}      -0.059118    1.345771    0.922933

\textbf{CH$_3$OH}\\
 C \tabto{1cm}     -0.046586    0.666185    0.000000\\
 H \tabto{1cm}     -1.087128    0.977597    0.000000\\
 H \tabto{1cm}      0.438190    1.073682    0.889458\\
 H \tabto{1cm}      0.438190    1.073682   -0.889458\\
 O \tabto{1cm}     -0.046586   -0.757229    0.000000\\
 H \tabto{1cm}      0.862947   -1.064240    0.000000

\textbf{CH}\\
 C  \tabto{1cm}     0.000000    0.000000    0.159626\\
 H  \tabto{1cm}     0.000000    0.000000   -0.957757

\textbf{VdW}\\
 C  \tabto{1cm}     -1.125250   -0.282677    0.060208\\
 H  \tabto{1cm}     -2.011990    0.342810   -0.002464\\
 H  \tabto{1cm}     -1.133654   -1.031295   -0.722851\\
 H  \tabto{1cm}     -1.039412   -0.756251    1.034640\\
 O  \tabto{1cm}      0.044095    0.535155   -0.189846\\
 H  \tabto{1cm}      0.091489    1.248675    0.456345\\
 C  \tabto{1cm}      1.520228   -0.225107    0.161872\\
 H  \tabto{1cm}       1.370941   -1.038472   -0.579380

\textbf{TS-insCH}\\
 C  \tabto{1cm}       0.393776    0.663697    0.008555\\
 H  \tabto{1cm}       0.731391    1.328107   -0.783364\\
 H  \tabto{1cm}      -0.789585    0.660933   -0.239915\\
 H  \tabto{1cm}       0.403111    1.138477    0.986879\\
 O  \tabto{1cm}       1.084088   -0.522138   -0.072860\\
 H  \tabto{1cm}       0.858968   -1.072930    0.684215\\
 C  \tabto{1cm}      -1.802399   -0.142893    0.068772\\
 H  \tabto{1cm}      -1.424852   -1.002302   -0.528893

\textbf{TS-insOH}\\
 C   \tabto{1cm}     -1.155761    0.251122    0.030032\\
 H   \tabto{1cm}     -1.349390    0.563207    1.052289\\
 H   \tabto{1cm}     -0.975332    1.128684   -0.589128\\
 H   \tabto{1cm}     -1.991161   -0.327839   -0.355035\\
 O   \tabto{1cm}      0.006792   -0.600015    0.032293\\
 H   \tabto{1cm}      0.615270   -0.331930   -0.863154\\
 C   \tabto{1cm}      1.429149    0.389996   -0.069677\\
 H   \tabto{1cm}      2.005950   -0.078704    0.734551

\textbf{RI1}\\
 C  \tabto{1cm}     -1.183629   -0.197282   -0.000141\\
 H  \tabto{1cm}     -1.279163   -0.852779    0.863160\\
 H  \tabto{1cm}     -1.309326   -0.794998   -0.901081\\
 H  \tabto{1cm}     -1.982255    0.543751    0.033803\\
 C  \tabto{1cm}      0.175766    0.482798   -0.003353\\
 H  \tabto{1cm}      0.282617    1.196707   -0.833505\\
 H  \tabto{1cm}      0.311806    1.104841    0.899271\\
 O  \tabto{1cm}      1.252937   -0.363827   -0.005086

\textbf{RI2}\\
 C  \tabto{1cm}       1.232920   -0.271107    0.008952\\
 H  \tabto{1cm}       2.125658    0.101853    0.485772\\
 H  \tabto{1cm}       1.288277   -1.185208   -0.562365\\
 C  \tabto{1cm}      -0.005441    0.539806   -0.032130\\
 H  \tabto{1cm}      -0.024020    1.254200    0.796673\\
 H  \tabto{1cm}      -0.062224    1.126278   -0.954800\\
 O  \tabto{1cm}      -1.192797   -0.256587   -0.041871\\
 H  \tabto{1cm}      -1.150188   -0.856623    0.708756

\textbf{RI3}\\
 C   \tabto{1cm}     1.225375   -0.164412    0.011796\\
 H   \tabto{1cm}     1.299638   -0.987729   -0.699616\\
 H   \tabto{1cm}     1.391315   -0.584669    1.011574\\
 H   \tabto{1cm}     2.027674    0.541578   -0.192456\\
 C   \tabto{1cm}    -0.088679    0.507311   -0.099182\\
 H   \tabto{1cm}    -0.233773    1.523984    0.243669\\
 O   \tabto{1cm}    -1.165758   -0.340238    0.019741\\
 H   \tabto{1cm}    -1.978962    0.171346    0.003216

\textbf{RI4}\\
 C  \tabto{1cm}    -1.198182   -0.227975    0.063672\\
 H  \tabto{1cm}    -2.122297    0.315789   -0.029786\\
 H  \tabto{1cm}    -1.121028   -1.272339   -0.207260\\
 O  \tabto{1cm}    -0.092240    0.546023   -0.035119\\
 C  \tabto{1cm}     1.135705   -0.170023    0.011955\\
 H  \tabto{1cm}     1.930104    0.555714   -0.126891\\
 H  \tabto{1cm}     1.251511   -0.664961    0.976907\\
 H  \tabto{1cm}     1.174490   -0.914393   -0.785784
 
\textbf{VdW-TS18}\\
 H  \tabto{1cm}     1.169798    1.231005    0.921012\\
 O  \tabto{1cm}    -2.183000   -0.000002   -0.000001\\
 H  \tabto{1cm}    -1.206630   -0.000002   -0.000029\\
 C  \tabto{1cm}     1.165940    0.665178    0.000002\\
 H  \tabto{1cm}     1.169843    1.231009   -0.921005\\
 H  \tabto{1cm}     1.169807   -1.231003    0.921011\\
 C  \tabto{1cm}     1.165948   -0.665176    0.000002\\
 H  \tabto{1cm}     1.169856   -1.231003   -0.921007

\textbf{TS1}\\
 C  \tabto{1cm}     -1.478127   -0.169193    0.000000\\
 H  \tabto{1cm}     -1.456637   -0.730762    0.919864\\
 H  \tabto{1cm}     -1.456643   -0.730747   -0.919873\\
 H  \tabto{1cm}     -2.001791    0.775335    0.000010\\
 C  \tabto{1cm}      0.545858    0.549378   -0.000000\\
 H  \tabto{1cm}      0.355888    1.107068   -0.930074\\
 H  \tabto{1cm}      0.355890    1.107069    0.930073\\
 O  \tabto{1cm}      1.224613   -0.476135    0.000000

\textbf{TS2}\\
 C   \tabto{1cm}       1.184162   -0.169425    0.037011\\
 H   \tabto{1cm}       1.508026   -0.679879   -0.872835\\
 H   \tabto{1cm}       1.208265   -0.877237    0.860709\\
 H   \tabto{1cm}       1.870825    0.654656    0.219463\\
 C   \tabto{1cm}      -0.222534    0.322669   -0.197155\\
 H   \tabto{1cm}      -0.293200    1.478203    1.101937\\
 H   \tabto{1cm}      -0.312051    1.219391   -0.840334\\
 O   \tabto{1cm}      -1.218954   -0.339325    0.061490

\textbf{TS3}\\
 C   \tabto{1cm}    -0.789756   -0.253889    0.000000\\
 H   \tabto{1cm}    -1.179469   -0.642131    0.926477\\
 H   \tabto{1cm}    -1.179469   -0.642132   -0.926477\\
 H   \tabto{1cm}    -1.991080    0.832530   -0.000001\\
 C   \tabto{1cm}     0.327718    0.714704    0.000000\\
 H   \tabto{1cm}     0.495441    1.278048   -0.913925\\
 H   \tabto{1cm}     0.495440    1.278048    0.913925\\
 O   \tabto{1cm}     0.766421   -0.608657   -0.000000

\textbf{TS4}\\
 C  \tabto{1cm}     -1.045239   -0.206114   -0.000000\\
 H  \tabto{1cm}     -1.591684   -0.332173    0.924563\\
 H  \tabto{1cm}     -0.076310   -1.128099    0.000001\\
 H  \tabto{1cm}     -1.591683   -0.332174   -0.924563\\
 C  \tabto{1cm}      0.146714    0.696861   -0.000000\\
 H  \tabto{1cm}      0.283974    1.297940   -0.895971\\
 H  \tabto{1cm}      0.283974    1.297940    0.895971\\
 O  \tabto{1cm}      1.010360   -0.468489    0.000000

\textbf{TS5}\\
 C   \tabto{1cm}     1.217381   -0.174502    0.001619\\
 H   \tabto{1cm}     1.456928   -0.473454   -1.023314\\
 H   \tabto{1cm}     1.229242   -1.068808    0.620614\\
 H   \tabto{1cm}     1.990458    0.512644    0.341637\\
 C   \tabto{1cm}    -0.124087    0.477632    0.028880\\
 H   \tabto{1cm}    -0.964732    0.245316    0.945733\\
 H   \tabto{1cm}    -0.217545    1.491504   -0.357724\\
 O   \tabto{1cm}    -1.256764   -0.315748   -0.088743

\textbf{TS6}\\
 C   \tabto{1cm}     1.251515   -0.219935    0.028208\\
 H   \tabto{1cm}     1.249303   -1.283119   -0.145135\\
 H   \tabto{1cm}     2.093880    0.249739    0.509762\\
 H   \tabto{1cm}     0.778408    0.380931   -1.043397\\
 C   \tabto{1cm}    -0.042117    0.478482   -0.041948\\
 H   \tabto{1cm}    -0.090139    1.490274    0.342952\\
 O   \tabto{1cm}    -1.170817   -0.316442    0.089645\\
 H   \tabto{1cm}    -1.921302    0.142430   -0.298902

\textbf{TS7}\\
 C  \tabto{1cm}     0.947498   -0.238200    0.000003\\
 H  \tabto{1cm}     1.391838   -0.559459   -0.926793\\
 H  \tabto{1cm}     1.391834   -0.559448    0.926805\\
 C  \tabto{1cm}    -0.090144    0.780696   -0.000005\\
 H  \tabto{1cm}    -0.264307    1.345069    0.909689\\
 H  \tabto{1cm}    -0.264303    1.345059   -0.909705\\
 O  \tabto{1cm}    -0.666427   -0.529376    0.000002\\
 H  \tabto{1cm}    -2.067769   -0.591188   -0.000001

\textbf{TS8}\\
 C   \tabto{1cm}    -1.223160   -0.237032    0.004239\\
 H   \tabto{1cm}    -2.122688    0.304707   -0.234788\\
 H   \tabto{1cm}    -1.298175   -1.221532    0.438495\\
 C   \tabto{1cm}    -0.029031    0.354124   -0.171706\\
 H   \tabto{1cm}     0.060806    1.295196   -0.697956\\
 H   \tabto{1cm}    -0.038731    1.427464    1.352852\\
 O   \tabto{1cm}     1.129433   -0.345797    0.017978\\
 H   \tabto{1cm}     1.876471    0.257994    0.002375

\textbf{TS10}\\
 C   \tabto{1cm}    -1.210861   -0.192441   -0.023519\\
 H   \tabto{1cm}    -1.991207    0.506753   -0.318420\\
 H   \tabto{1cm}    -1.460662   -0.561814    0.977711\\
 H   \tabto{1cm}    -1.180838   -1.041638   -0.700732\\
 C   \tabto{1cm}     0.117972    0.476636    0.049201\\
 H   \tabto{1cm}     0.144765    1.519096    0.393444\\
 O   \tabto{1cm}     1.183361   -0.114277   -0.159097\\
 H   \tabto{1cm}     1.578389   -1.213347    0.766679

\textbf{TS11}\\
 C   \tabto{1cm}     -1.149882    0.076617   -0.219041\\
 H   \tabto{1cm}     -1.260377    1.092781   -0.563605\\
 H   \tabto{1cm}     -1.777518    0.797486    1.610272\\
 H   \tabto{1cm}     -2.012239   -0.567895   -0.213649\\
 C   \tabto{1cm}      0.055346   -0.440309    0.030154\\
 H   \tabto{1cm}      0.189051   -1.472286    0.327499\\
 O   \tabto{1cm}      1.185292    0.317622   -0.031517\\
 H   \tabto{1cm}      1.945970   -0.208905    0.224938

\textbf{TS12}\\
 C  \tabto{1cm}   -1.015093   -0.202738   -0.000000\\
 H  \tabto{1cm}   -1.207645   -0.774638   -0.902942\\
 H  \tabto{1cm}   -1.644865    0.691117   -0.000002\\
 H  \tabto{1cm}   -1.207645   -0.774635    0.902943\\
 C  \tabto{1cm}    0.806262   -0.468054   -0.000000\\
 H  \tabto{1cm}    1.213089   -0.860583    0.929501\\
 H  \tabto{1cm}    1.213091   -0.860582   -0.929501\\
 O  \tabto{1cm}    0.360870    0.825510    0.000000

\textbf{TS13}\\
 C   \tabto{1cm}    -1.236658    0.456106    0.000000\\
 H   \tabto{1cm}    -1.788766    0.336696    0.936812\\
 H   \tabto{1cm}    -1.788766    0.336696   -0.936812\\
 O   \tabto{1cm}     0.000000    0.636890    0.000000\\
 C   \tabto{1cm}     1.182753   -0.897101    0.000000\\
 H   \tabto{1cm}     2.123043   -0.371636    0.000000\\
 H   \tabto{1cm}     0.888959   -1.375454    0.920692\\
 H   \tabto{1cm}     0.888959   -1.375454   -0.920692

\textbf{TS14}\\
 C   \tabto{1cm}   -0.877348   -0.308985   -0.000000\\
 H   \tabto{1cm}   -1.393594   -0.519039    0.928777\\
 H   \tabto{1cm}   -1.393600   -0.519035   -0.928775\\
 O   \tabto{1cm}   -0.053399    0.818165    0.000001\\
 C   \tabto{1cm}    0.733600   -0.385309   -0.000001\\
 H   \tabto{1cm}    2.034461    0.337757    0.000002\\
 H   \tabto{1cm}    1.021205   -0.839620    0.940372\\
 H   \tabto{1cm}    1.021206   -0.839616   -0.940375

\textbf{TS15}\\
 C  \tabto{1cm}      1.162722   -0.276498    0.072234\\
 H  \tabto{1cm}      1.423240   -1.012556   -0.683791\\
 H  \tabto{1cm}     -0.245606   -0.820919    0.429523\\
 H  \tabto{1cm}      1.986398    0.034277    0.703301\\
 C  \tabto{1cm}      0.165922    0.667880   -0.233278\\
 H  \tabto{1cm}     -0.006543    1.579211    0.323325\\
 O  \tabto{1cm}     -1.173037   -0.277139   -0.055467\\
 H  \tabto{1cm}     -1.745057    0.088806    0.637639

\textbf{TS17}\\
 C   \tabto{1cm}     1.110849   -0.515704    0.129558\\
 H   \tabto{1cm}     2.012593   -0.761255   -0.413380\\
 H   \tabto{1cm}    -0.265282   -0.955452    0.105463\\
 C   \tabto{1cm}     0.412754    0.693506    0.006471\\
 H   \tabto{1cm}     0.154471    1.283353    0.881168\\
 H   \tabto{1cm}     0.378537    1.253260   -0.924550\\
 O   \tabto{1cm}    -1.204482   -0.224403   -0.136702\\
 H   \tabto{1cm}    -1.786078   -0.091490    0.628743

\textbf{TS18}\\
 H  \tabto{1cm}      -1.171596   -1.197183    0.883063\\
 O  \tabto{1cm}       1.989297   -0.078392    0.004837\\
 H  \tabto{1cm}       1.347281   -0.603871   -0.501622\\
 C  \tabto{1cm}      -1.296520   -0.551832    0.024952\\
 H  \tabto{1cm}      -1.801606   -0.970265   -0.834463\\
 H  \tabto{1cm}      -0.363245    1.126150    0.894760\\
 C  \tabto{1cm}      -0.859758    0.705005    0.032152\\
 H  \tabto{1cm}      -0.987539    1.353267   -0.823058

\textbf{TS19}\\
 C   \tabto{1cm}    -1.161215   -0.276272    -0.016104\\
 H   \tabto{1cm}    -2.044642    0.017635    0.527912\\
 H   \tabto{1cm}    -1.183930   -1.167256   -0.627378\\
 C   \tabto{1cm}    -0.056163    0.670436   -0.170115\\
 H   \tabto{1cm}     0.773291    1.381044   -0.285714\\
 H   \tabto{1cm}    -0.178921    1.097466    0.867021\\
 O   \tabto{1cm}     1.134970   -0.336845   -0.001140\\
 H   \tabto{1cm}     0.858710   -0.999119    0.644587
 
\textbf{CH$_3$}\\
 C    \tabto{1cm}     -0.000001    0.000000    0.000001\\
 H    \tabto{1cm}      0.537577    0.932074   -0.000002\\
 H    \tabto{1cm}      0.538418   -0.931589   -0.000002\\
 H    \tabto{1cm}     -1.075990   -0.000485   -0.000002

\textbf{H$_2$CO}\\
 C    \tabto{1cm}      -0.000000    0.530112    0.000000\\
 H    \tabto{1cm}       0.000000    1.111691    0.936283\\
 H    \tabto{1cm}       0.000000    1.111691   -0.936283\\
 O    \tabto{1cm}      -0.000000   -0.675507   -0.000000

\textbf{ep-CH$_2$CH$_2$O}\\
 C     \tabto{1cm}       0.000000   -0.821596    0.000000\\
 H     \tabto{1cm}       0.052950   -1.393070    0.917610\\
 H     \tabto{1cm}       0.052950   -1.393070   -0.917610\\
 C     \tabto{1cm}      -0.667590    0.478899    0.000000\\
 H     \tabto{1cm}      -1.101076    0.855031   -0.917610\\
 H     \tabto{1cm}      -1.101076    0.855031    0.917610\\
 O     \tabto{1cm}       0.762724    0.391533    0.000000

\textbf{CH$_3$CHO}\\
 C     \tabto{1cm}       0.000000    0.460239    0.000000\\
 H     \tabto{1cm}       0.486731    1.454146   -0.000000\\
 O     \tabto{1cm}      -1.206196    0.376321    0.000000\\
 C     \tabto{1cm}       0.935294   -0.711258   -0.000000\\
 H     \tabto{1cm}       1.583762   -0.656041    0.876115\\
 H     \tabto{1cm}       0.383550   -1.646524   -0.000000\\
 H     \tabto{1cm}       1.583762   -0.656041   -0.876115

\textbf{CH$_2$CH$_2$}\\
 C   \tabto{1cm}     0.000000    0.664070    0.000000\\
 H   \tabto{1cm}     0.920941    1.229754    0.000000\\
 H   \tabto{1cm}    -0.920940    1.229756    0.000000\\
 C   \tabto{1cm}    -0.000000   -0.664070    0.000000\\
 H   \tabto{1cm}    -0.920941   -1.229754    0.000000\\
 H   \tabto{1cm}     0.920940   -1.229756    0.000000

\textbf{OH}\\
 O    \tabto{1cm}     0.000000    0.000000    0.108015\\
 H    \tabto{1cm}     0.000000    0.000000   -0.864117

\textbf{H$_2$CCHOH}\\
 C     \tabto{1cm}    1.235153   -0.069728    0.000000\\
 H     \tabto{1cm}    2.071302    0.609113    0.000000\\
 H     \tabto{1cm}    1.421615   -1.132691    0.000000\\
 C     \tabto{1cm}    0.000000    0.415973    0.000000\\
 H     \tabto{1cm}   -0.200321    1.479988    0.000000\\
 O     \tabto{1cm}   -1.100807   -0.396584    0.000000\\
 H     \tabto{1cm}   -1.897055    0.138786    0.000000

\textbf{H$_2$CCH}\\
 C  \tabto{1cm}      0.048619   -0.585488   -0.000000\\
 H  \tabto{1cm}     -0.880007   -1.154363   -0.000000\\
 H   \tabto{1cm}      0.965820   -1.162024    0.000000\\
 C   \tabto{1cm}      0.048619    0.718179    0.000000\\
 H   \tabto{1cm}     -0.669237    1.520237    0.000000

\textbf{H$_2$O}\\
 O    \tabto{1cm}     0.000000   -0.000000    0.117275\\
 H    \tabto{1cm}    -0.000000    0.761078   -0.469101\\
 H    \tabto{1cm}    -0.000000   -0.761078   -0.469101

\section{Energies}

\begin{table*}
	\centering
	\caption{Electronic and ZPE-corrected energies of all the involved optimized compounds (in Hartrees), at the B2PLYP-D3/aug-cc-pVTZ level of theory; CCSD(T)/aug-cc-pVTZ electronic energies reevaluations and corrections with the ZPE obtained with the previous method.}
	\label{tab:elec-energies}
	\begin{tabular}{lccccc} 
		\hline
		\textbf{Compound} & \multicolumn{3}{c}{\textbf{B2PLYP-D3}} & \multicolumn{2}{c}{\textbf{CCSD(T)}}\\
		 & \textit{Elec.} & \textit{ZPE} & \textit{ZPE-corr} & \textit{Elec.} & \textit{ZPE-corr} \\
		\hline
		\textbf{C$_2$H$_5$} & -79.10668879 & 0.059671789 & -79.047017 & -79.0079052 & -78.94823341\\
		\textbf{O($^3$P)} & -75.05099966 & 0 & -75.05099966 & -74.9789523 & -74.9789523\\
		\textbf{CH$_3$OH} & -115.6848316 & 0.051493616 & -115.633338 & -115.5623529 & -115.5108593\\
		\textbf{CH} & -38.45881718 & 0.006566177 & -38.452251 & -38.4128059 & -38.40623972\\
		\textbf{VdW} & -154.1750537 & 0.064407749 & -154.110646 & -154.0031607 & -153.938753\\
		\textbf{TS-insCH} & -154.1545473 & 0.059343288 & -154.095204 & -153.9820834 & -153.9227401\\
		\textbf{TS-insOH} & -154.1585606 & 0.059288608 & -154.099272 & -153.985481 & -153.9261924\\
		\textbf{RI1} & -154.3060805 & 0.065237505 & -154.240843 & -154.1318969 & -154.0666594\\
		\textbf{RI2} & -154.3101518 & 0.065377849 & -154.244774 & -154.13636 & -154.0709822\\
		\textbf{RI3} & -154.3230296 & 0.066298595 & -154.256731 & -154.1480104 & -154.0817118\\
		\textbf{RI4} & -154.3021616 & 0.066081578 & -154.23608 & -154.1268493 & -154.0607677\\
		\textbf{VdW-TS18} & -154.2672303 & 0.061910314 & -154.20532 & -154.0943931 & -154.0324828\\
        \textbf{TS1} & -154.2760087 & 0.061650654 & -154.214358 & -154.0999523 & -154.0383016\\
        \textbf{TS2} & -154.2646351 & 0.057734147 & -154.206901 & -154.0887988 & -154.0310647\\
        \textbf{TS3} & -154.2136588 & 0.05972579 & -154.153933 & -154.0388684 & -153.9791426\\
        \textbf{TS4} & -154.2569693 & 0.062110339 & -154.194859 & -154.0826561 & -154.0205458\\
        \textbf{TS5} & -154.2586956 & 0.061406608 & -154.197289 & -154.0837379 & -154.0223313\\
        \textbf{TS6} & -154.2458148 & 0.061430801 & -154.184384 & -154.0705995 & -154.0091687\\
        \textbf{TS7} & -154.2081293 & 0.058303274 & -154.149826 & -154.0315838 & -153.9732805\\
        \textbf{TS8} & -154.2503579 & 0.058349873 & -154.192008 & -154.0747988 & -154.0164489\\
        \textbf{TS10} & -154.2599801 & 0.057030093 & -154.20295 & -154.0820002 & -154.0249701\\
        \textbf{TS11} & -154.2546143 & 0.057931297 & -154.196683 & -154.0784341 & -154.0205028\\
        \textbf{TS12} & -154.2214067 & 0.063331708 & -154.158075 & -154.0467308 & -153.9833991\\
        \textbf{TS13} & -154.2569183 & 0.060760265 & -154.196158 & -154.0795803 & -154.01882\\
        \textbf{TS14} & -154.1996478 & 0.059287811 & -154.140360 & -154.0245560 & -153.9652682\\
        \textbf{TS15} & -154.1983588 & 0.058515762 & -154.139843 & -154.0218061 & -153.9632903\\
        \textbf{TS17} & -154.1906219 & 0.057473896 & -154.133148 & -154.0136025 & -153.9561286\\
        \textbf{TS18} & -154.2661948 & 0.060923789 & -154.205271 & -154.0930633 & -154.0321395\\
        \textbf{TS19} & -154.2116246 & 0.061062645 & -154.150562 & -154.0374376 & -153.976375\\
        \textbf{CH$_3$} & -39.81255894 & 0.029963943 & -39.782595 & -39.7636363 & -39.73367236\\
        \textbf{H$_2$CO} & -114.4688577 & 0.026675717 & -114.442182 & -114.34288 & -114.3162043\\
        \textbf{ep-CH$_2$CH$_2$O} & -153.7306823 & 0.05767932 & -153.673003 & -153.5565781 & -153.4988988\\
        \textbf{H} & -0.498668238 & 0 & -0.498668238 & -0.499821176 & -0.499821176\\
        \textbf{CH$_3$CHO} & -153.7743088 & 0.055628812 & -153.71868 & -153.5983913 & -153.5427625\\
        \textbf{CH$_2$CH$_2$} & -78.54273093 & 0.05117193 & -78.491559 & -78.4436342 & -78.39246227\\
        \textbf{OH} & -75.7193689 & 0.008511898 & -75.710857 & -75.6455848 & -75.6370729\\
        \textbf{H$_2$CCHOH} & -153.7563476 & 0.056364564 & -153.699983 & -153.5807646 & -153.5244\\
        \textbf{H$_2$CCH} & -77.85799294 & 0.036751943 & -77.821241 & -77.7569861 & -77.72023416\\
        \textbf{H$_2$O} & -76.41484596 & 0.021349957 & -76.393496 & -76.3423167 & -76.32096674\\
		\hline
	\end{tabular}
\end{table*}


\bsp	
\label{lastpage}
\end{document}